\documentclass{article}
\DeclareUnicodeCharacter{2212}{-}
\DeclareUnicodeCharacter{22C5}{\ensuremath{\cdot}}
\usepackage{subcaption}
\usepackage[english]{babel}

\usepackage[letterpaper,top=2cm,bottom=2cm,left=3cm,right=3cm,marginparwidth=1.75cm]{geometry}

\usepackage{amsmath}
\usepackage{graphicx}
\usepackage[colorlinks=true, allcolors=blue]{hyperref}
\usepackage{authblk}
\usepackage{subcaption}
\usepackage{float}
\usepackage{tabularx}
\usepackage{array}

\title{Peptide Structure Prediction Using Counter-Diabatic Quantum Approximate Optimization Algorithm (CD-QAOA)}

\author[1]{Sung Won Yun$^{\dagger}$}
\author[2]{Yeon Gyo Seo$^{\dagger}$}
\author[3]{Seong Hun Jang}
\author[3]{Suhyun Park}
\author[1]{Joonwoo Bae$^{*}$}
\author[2,3,4]{Sangwook Wu$^{*}$}
\affil[1]{School of Electrical Engineering, Korea Advanced Institute of Science and Technology (KAIST), 291 Daehak-ro,
Yuseong-gu, Daejeon 34141, Republic of Korea}
\affil[2]{Department of Physics, Pukyong National University, Yongso-ro, Busan 48513, Republic of Korea}
\affil[3]{PharmCADD, 1102-ho, 60, Centum Jungang-ro, Haeundae-gu, Busan 48059, Republic of Korea}
\affil[4]{Quantum AI Institute of Pukyong National University, Yongso-ro, Pukyong National University, Busan 48513, Republic of Korea }
\date{}
\begin{document}
\maketitle

\begin{abstract}
\noindent In this study, we predicted the structure of the heptapeptide APRLRFY, a neuropeptide sequence, on a tetrahedral lattice using a Quantum Approximate Optimization Algorithm (QAOA). QAOA is a gate-based quantum heuristic designed to obtain approximate solutions to combinatorial optimization problems by mapping a Quadratic Unconstrained Binary Optimization (QUBO) problem onto a parameterized quantum circuit. QAOA is based on the adiabatic approximation and has been successfully applied to a wide range of optimization problems. However, relatively slow convergence during ground-state searches has frequently been reported. To overcome this limitation, we employed the Counter-Diabatic Quantum Approximate Optimization Algorithm (CD-QAOA), which introduces an additional counter-diabatic driving term into the adiabatic framework to accelerate convergence toward the ground state during peptide structure prediction. In the heptapeptide structure prediction, intermolecular interactions were modeled using two different approaches. In the first approach, only the interaction between the Second residue, proline (P), and the seventh residue, tyrosine (Y), was included in the optimization. In the second approach, all residue–residue interactions within the heptapeptide were modeled using the Miyazawa–Jernigan (MJ) interaction matrix.
To validate the peptide structures predicted using CD-QAOA, we additionally employed several classical computational methods, including quantum chemistry–based Hartree–Fock (HF) calculation and Density Functional Theory (DFT) calculation, conventional molecular dynamics (MD) simulation, and Hamiltonian replica exchange molecular dynamics (H-REMD) simulation. The structural similarities among the conformations obtained from these different approaches were systematically analyzed. CD-QAOA is highly effective for predicting the structures of short peptides. In particular, we demonstrate that a quantum–classical hybrid framework can significantly improve both the efficiency and accuracy of peptide structure prediction.

\end{abstract}
\let\thefootnote\relax\footnotetext{$\dagger$ These authors contributed equally to this work.}
\let\thefootnote\relax\footnotetext{$^{*}$Corresponding author. Email: \texttt{joonwoo.bae@kaist.ac.kr, s.wu@pharmcadd.com}}
\section{Introduction}

Since the 1960s, following the formulation of the Levinthal paradox, determining the three-dimensional structure of proteins from their amino acid sequences has become one of the most challenging problems in physics, chemistry, and the life sciences \cite{Levinthal1968,Levinthal1969,Karplus1997,Ivankov2020}. With the recent revolution in artificial intelligence, AlphaFold \cite{Jumper2021} has achieved a groundbreaking milestone in protein structure prediction. One of the key factors behind the success of AI-based protein structure prediction is the availability of the Protein Data Bank (PDB) \cite{Burley2024}, a repository that stores three-dimensional structural information of biomacromolecules such as proteins, DNA, and RNA. Established in 1971, the PDB has served as a central public database essential for drug discovery, biotechnology, and fundamental biological research. As of 2026, the PDB contains more than 250,000 experimentally determined biomolecular structures \cite{PDBe2026}. Thousands of new structures determined by techniques such as X-ray crystallography, NMR spectroscopy, and cryo-electron microscopy (cryo-EM) are deposited annually.\\

\begin{figure}[htbp]
\centering
\includegraphics[width=\linewidth]{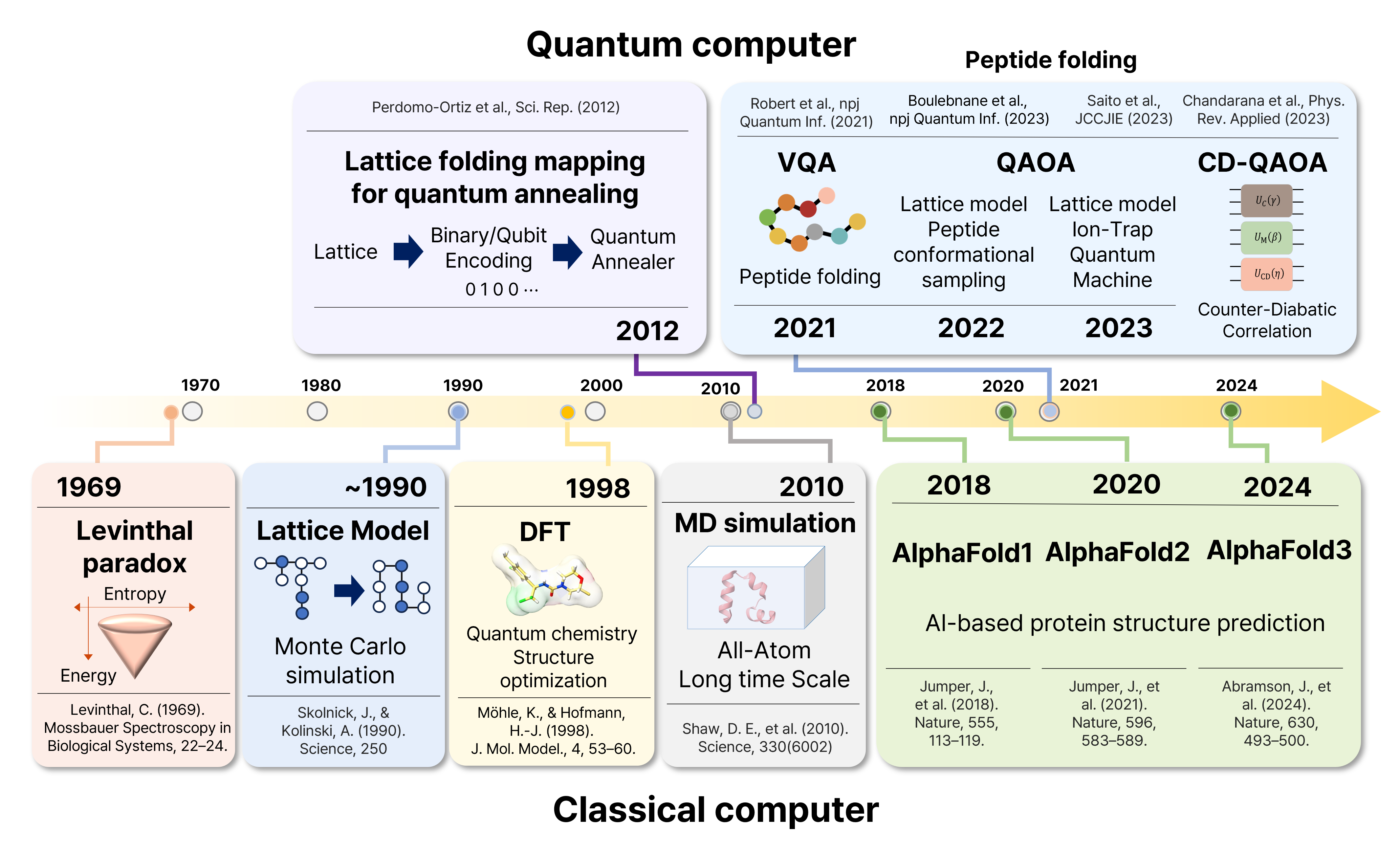}
\caption{Historical overview of protein-folding approaches based on classical and quantum computing.}
\label{fig:history}
\end{figure}

However, despite the remarkable success of AlphaFold, significant limitations remain in accurately predicting the flexibility and dynamic behavior of short peptides, particularly those consisting of fewer than 20 amino acids, due to the limited availability of peptide structural data in existing databases \cite{Singh2016,Wang2016,Wang2015}. Existing peptide databases contain significantly fewer experimentally resolved structures for short and highly flexible peptides compared with globular proteins.\\

Since the late 2010s, there has been growing interest in addressing protein folding---particularly peptide folding---by mapping the problem onto mathematical optimization frameworks defined on lattice models and solving them using quantum algorithms \cite{kumar,Boulebnane,Robert,Doga}. These approaches bear similarities to earlier efforts prior to the advent of GPUs, in which proteins were simplified into hydrophobic and hydrophilic components and their ground states were explored using Monte Carlo simulations \cite{Lau1989} on classical CPUs as shown in Fig.~\ref{fig:history}.\\

Among various quantum algorithms, the Variational Quantum Eigensolver (VQE) \cite{Peruzzo2014,McClean2016} and QAOA \cite{Farhi2014,Farhi2014b,Preskill2018} are representative methods widely used for optimization tasks. In particular, QAOA is a prominent variational quantum algorithm designed to solve combinatorial optimization problems within a circuit-based framework, with broad applicability across diverse domains. QAOA is considered particularly suitable for the current Noisy Intermediate-Scale Quantum (NISQ) era. It employs a hybrid quantum--classical architecture, where the quantum processor handles state preparation and measurement, while a classical optimizer updates the variational parameters ($\beta$, $\gamma$). This hybrid structure enables the efficient utilization of limited quantum resources. Moreover, its relatively shallow circuit depth ($p$) helps mitigate hardware errors. By reformulating problems into the QUBO framework, QAOA can be flexibly applied to a wide range of combinatorial optimization problems, including Max-Cut \cite{Lucas2014}, portfolio optimization in finance \cite{Brandhofer2023}, vehicle routing problems (VRP) \cite{Azad2022}. Despite these advantages, several limitations remain in practical applications. A key factor determining the performance of QAOA is its convergence behavior, which describes the probabilistic and mathematical pathway toward reaching the optimal solution as the number of layers ($p$) increases. From a theoretical perspective, QAOA can be viewed as a discrete approximation of adiabatic quantum computing, and its convergence is guaranteed in the limit of large $p$. Through Trotter decomposition \cite{Trotter1959}, the sequence of unitary operations approximates adiabatic state evolution, ensuring that the system remains in the ground state and converges to the optimal solution with a probability approaching unity.\\

However, in practice, QAOA introduces variational parameters ($\beta$, $\gamma$) that must be optimized classically. As the parameter space grows, the optimization landscape may suffer from the barren plateau phenomenon \cite{McClean2018}, where gradients vanish and convergence slows dramatically, especially for large systems. While increasing $p$ theoretically improves accuracy, in real hardware environments, accumulated noise can degrade performance, leading to a trade-off between circuit depth and solution quality.\\

To address these limitations, the Counter-Diabatic Quantum Approximate Optimization Algorithm (CD-QAOA) \cite{Wurtz2021,Vizzuso2024,Berry2009,GueryOdelin2019} has been proposed as an advanced variational quantum algorithm. It aims to accelerate adiabatic state preparation within finite circuit depth by incorporating the concept of a counter-diabatic driving term, a strategy from shortcuts to adiabaticity. This approach suppresses diabatic transitions that occur during rapid system evolution, enabling the system to reach the ground state with high fidelity even at low circuit depths. CD-QAOA introduces an additional correction term, known as the counter-diabatic gauge potential \cite{demirplak2003}. In practice, instead of computing the exact counter-diabatic term, a variational approximation---referred to as variational counter-diabatic driving---is employed to construct the correction operator. This often involves adding commutator terms to the quantum circuit, thereby improving optimization efficiency. A key advantage of CD-QAOA is its ability to accelerate convergence. The corrected driving term helps prevent transitions to excited states, particularly in regions where the energy gap becomes small. As a result, even for complex problems where standard QAOA requires large $p$, CD-QAOA can achieve high approximation ratios with shallow circuits, such as $p = 1$ or $2$. This provides a practical pathway toward achieving quantum advantage while minimizing gate errors. Furthermore, although CD-QAOA increases the number of variational parameters, it smooths the energy landscape, reducing the likelihood of getting trapped in local minima and improving robustness against initial parameter selection.\\

The $\alpha$-Bag Cell Peptide ($\alpha$-BCP) is a neuropeptide that plays a key role in regulating reproductive behavior in the mollusk Aplysia californica \cite{Rothman1983,Pulst1987}. It is produced and released, along with Egg-Laying Hormone (ELH) \cite{Pulst1986}, from a common precursor during the electrical discharge of bag cells located in the abdominal ganglion \cite{Rock1986}. The heptapeptide sequence APRLRFY serves as a core motif responsible for the biological activity of $\alpha$-BCP. It functions as a local neuromodulator by binding to G protein-coupled receptors on neighboring neurons, thereby regulating neuronal excitability.\\

In this study, we predict the three-dimensional structure of a heptapeptide on a tetrahedral lattice using the CD-QAOA framework. In addition to the quantum algorithm approach, three complementary methods are employed for comparison. First, quantum chemical calculations are performed to determine the most stable molecular structure based on Hartree-Fock (HF) and Density Functional Theory (DFT). Second, the AI-predicted structure is subjected to molecular dynamics simulations in an explicit solvent environment to obtain an energy-minimized conformation. Third, the conformational space of the heptapeptide is explored using the Hamiltonian Replica Exchange Molecular Dynamics (H-REMD) method, which is widely used for studying protein folding, and candidate structures are identified from the resulting energy landscape. A comparative analysis of these approaches is then conducted to elucidate the structural characteristics of the heptapeptide.

\subsection{Quantum computing approach  protein structure prediction}
In this subsection, we approach the problem of protein structure prediction using quantum algorithms. The quantum algorithm begins with the construction of a Hamiltonian such that its minimization provides a ground state from which a protein structure can be obtained.
We present a brief review of the construction of a Hamiltonian for this purpose, show how to realize a quantum circuit to find its minimum, and apply it to an amino acid sequence using about $10$ qubits.\\

Before constructing the Hamiltonian, we describe the encoding of the protein structure with qubits \cite{Robert}. For simplification, we assume that the $C_{\alpha}$ atoms of amino acids are located on a tetrahedral lattice. In the lattice, two types of nodes, A and B, are arranged alternately, with the directions of node B $(\bar{0},\bar{1},\bar{2},\bar{3})$ opposite to the corresponding directions of node A $(0,1,2,3)$; the bar indicates an outgoing direction from the node. Note that we place node B first; therefore, the sequence is $B\,A\,B\,A\cdots$. We encode the edge directions 0, 1, 2, and 3 as 00, 01, 10, and 11, respectively, which can be represented using two bits $q_i q_{i+1}$. The sequence of all edges is written as
\begin{equation}
[q_0q_1][q_2q_3][q_4q_5] \cdots[q_{2i-2}q_{2i-1}]\cdots [q_{2N-4}q_{2N-3}] \nonumber
\end{equation}
where $N$ is the number of residues. Note that we do not need the bar notation ($\bar{a}$) in the sequence. Whether the direction is outgoing or incoming can be inferred from whether the position is odd- or even-numbered.\

\subsubsection{Hamiltonian with the growth constraint}

Once a random even-length sequence is given, for example, \(0010011101 \cdots \), it can be mapped onto a protein backbone structure on the lattice. However, not every sequence corresponds to a valid structure. One invalid case occurs when the sequence contains repeated patterns among 00, 01, 10, and 11. For example, if the sequence contains a repetition such as $\cdots 10 \ 10 \cdots$, the structure first grows in the $10(\bar{10})$ direction and is then followed by growth in the $\bar{10}(10)$ direction, resulting in a conflict. To prevent this conflict, we introduce the growth constraint Hamiltonian \(H_{gc}\), defined as\\
\begin{equation}
H_{gc} = \sum^{N-1}_{i=1} \lambda_{gc} T(i)
\end{equation}

Note that $i$ can denote either the index of an amino acid $(1\leq i \leq N-1)$ or the index of an edge $(1\leq i \leq N-1)$. \(T(i)\) is equal to $1$ if the \(i\)th and \((i+1)\)th edges conflict; otherwise, it is equal to $0$. \(H_{gc}\) imposes a penalty by adding \(\lambda_{gc}\) for each pair of overlapping edge directions. To mathematically define \(T(i)\), we introduce the function $f_a(i)$, where $a \in \{0,1,2,3\}$.
\begin{align}
    f_0(i) &= (1-q_{2i-2})(1-q_{2i-1}) \\ 
    f_1(i) &= q_{2i-1}(q_{2i-1}-q_{2i-2}) \\
    f_2(i) &= q_{2i-2}(q_{2i-2}-q_{2i-1}) \\
    f_3(i) &= q_{2i-2}q_{2i-1}
\end{align}

If the direction of the $i$th node is $a$, then \(f_a(i)\) is equal to \(1\); otherwise, it is equal to 0. Thus, \(f_a(i)f_a(i+1)\) is equal to $1$ only when the \(i\)th and \((i+1)\)th edges are both equal to \(a\). Therefore, \(T(i)\) acts as an indicator of whether the \(i\)th and \((i+1)\)th edges are the same.\\
\begin{equation}
T(i)=\sum_{a \in \{0,1,2,3\}} f_a(i)f_a(i+1)
\end{equation}

\subsubsection{Hamiltonian for interactions}

Among all amino acids, interactions arise from van der Waals forces and Coulomb forces according to the electric charges of the side chains. These interactions determine the folded conformation. To represent the protein structure more accurately, we require not only $H_{gc}$ but also the interaction Hamiltonian $H_{int}$. Since interactions depend on distance, we first need $d(i,j)$, which is related to distance, before defining $H_{int}$. For convenience, $d(i,j)$ is not defined as the Euclidean distance. We also need the four-dimensional vector $\mathbf{x}(i,j)$ to define $d(i,j)$, which is given by
\begin{equation}
\mathbf{x}(i,j) = 
\begin{pmatrix}
\Delta n_0(i,j)\\ 
\Delta n_1(i,j) \\ 
\Delta n_2(i,j) \\ 
\Delta n_3(i,j) 
\end{pmatrix}
\end{equation}

where $\Delta n_a(i,j)$ is the effective number of edge directions $a$ between the $i$th and $j$th amino acids. As mentioned above, lattices A and B are arranged alternately. Also, A and B have opposite vector directions. Therefore, to determine how effectively the $i$th amino acid progresses to the $j$th amino acid in the $a$ direction, we count them while alternating the sign.\\
\begin{equation}
\Delta n_a(i,j) = \sum^{j-1}_{k=i} (-1)^i f_a(k)
\end{equation}

Using $\mathbf{x}(i,j)$, we obtain $d(i,j) = \Vert \mathbf{x}(i,j)\Vert^2$. $d(i,j)$ is set to $1$ only if the $i$th amino acid moves once in any direction to reach the $j$th amino acid. In other words, these $i$th and $j$th amino acids are at the closest distance. The next possible value of $d(i,j)$ is $2$, and $d(i,j)$ is equal to $2$ when the $i$th amino acid moves once each in any two directions. In other words, these $i$th and $j$th amino acids are at the second closest distance.\\

Now we can define $H_{int}$. Suppose the $i$th and $j$th amino acids are subject to an attractive interaction $\epsilon_{ij}$. $h_{ij}$ is the interaction Hamiltonian between the $i$th and $j$th amino acids.
\begin{equation}\label{1st hij}
h_{ij} = \epsilon_{ij}
\end{equation}

Building on Eq.~(\ref{1st hij}), we impose an additional penalty $\lambda_1$ when two nodes are not separated by the minimum distance, i.e., $d(i,j)\neq1$, thereby encouraging them to remain close.

\begin{equation}
h_{ij} = \epsilon_{ij} + \lambda_1 (d(i,j)-1)
\end{equation}

When they are at the closest distance, another type of conflict can occur that is not covered by $H_{gc}$. It only covers conflicts between adjacent amino acids. However, an amino acid can grow further and then change direction, possibly colliding with itself again. 
We define the set of amino acids adjacent to the $i$th amino acid as $\mathcal{N}(i)$. If one amino acid in $\mathcal{N}(i)$ grows to the $j$th amino acid, a collision occurs. Likewise, if one amino acid in $\mathcal{N}(j)$ grows to the $i$th amino acid, a conflict occurs. To prevent these conflicts, we force them to be at distance $d=2$ by introducing another penalty $\lambda_2$.
\begin{equation}
h_{ij} = \epsilon_{ij} + \lambda_1 (d(i,j)-1) 
+ \sum_{r\in\mathcal{N}(j)} \lambda_2 (2-d(i,r))
+ \sum_{m\in\mathcal{N}(i)} \lambda_2 (2-d(m,j))
\end{equation}

Note that the case $d(i,j)=1$ makes $d(\mathcal{N}(i),j)$ equal to $0$ or $2$, preventing the $\lambda_2$ term from acting as a negative penalty in the contact case. Furthermore, a large $\lambda_1$ prevents the negative penalty from being applied when there is no contact. Finally, we adopt $q_{ij}$ as
\begin{equation}
h_{ij} = q_{ij} \left(\epsilon_{ij} + \lambda_1 (d(i,j)-1) 
+ \sum_{r\in\mathcal{N}(j)} \lambda_2 (2-d(i,r))
+ \sum_{m\in\mathcal{N}(i)} \lambda_2 (2-d(m,j))\right).
\end{equation}

The activation of $q_{ij}$, where $q_{ij}=1$, is controlled by the classical optimizer. If $q_{ij}=1$ is selected, $h_{ij}$ is added. If a serious penalty occurs in the non-contact case, the classical optimizer sets $q_{ij}=0$ and turns it off. Finally, we sum $h_{i,j}$ to obtain $H_{int}$.
\begin{equation}
H_{int}=\sum_{i=1}^{N-4} \sum_{\substack{j \geq i+5 \\ j - i = 1 \mod 2}} h_{i,j}
\end{equation}

Here, a new condition is introduced such that $j-i$ is an odd number to prevent $d(i,j)=0$ beforehand.\\

\begin{figure}[t]
    \centering

    \begin{subfigure}{0.6\textwidth}
        \centering
        \includegraphics[width=\linewidth]{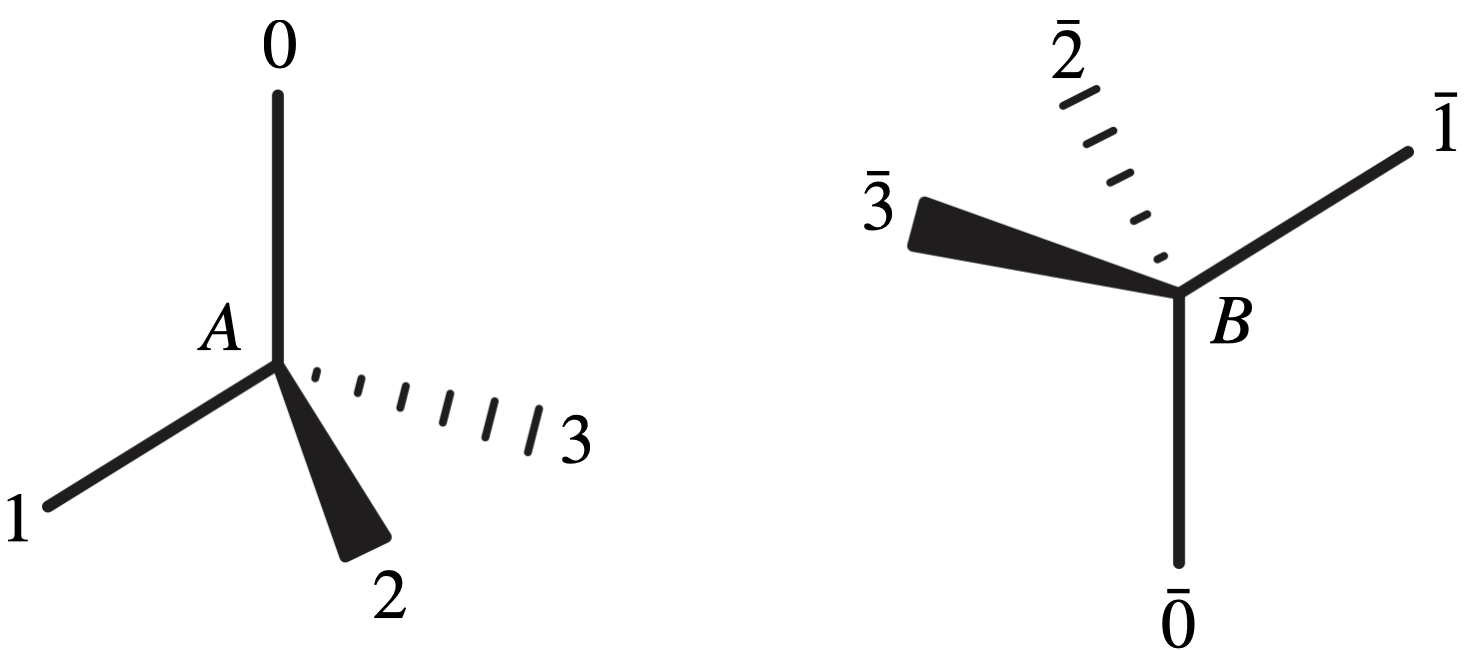}
        \caption{Tetrahedra A and B.}
        \label{fig:lattice}
    \end{subfigure}

    \vspace{0.5cm}

    \begin{subfigure}{0.6\textwidth}
        \centering
        \includegraphics[width=\linewidth]{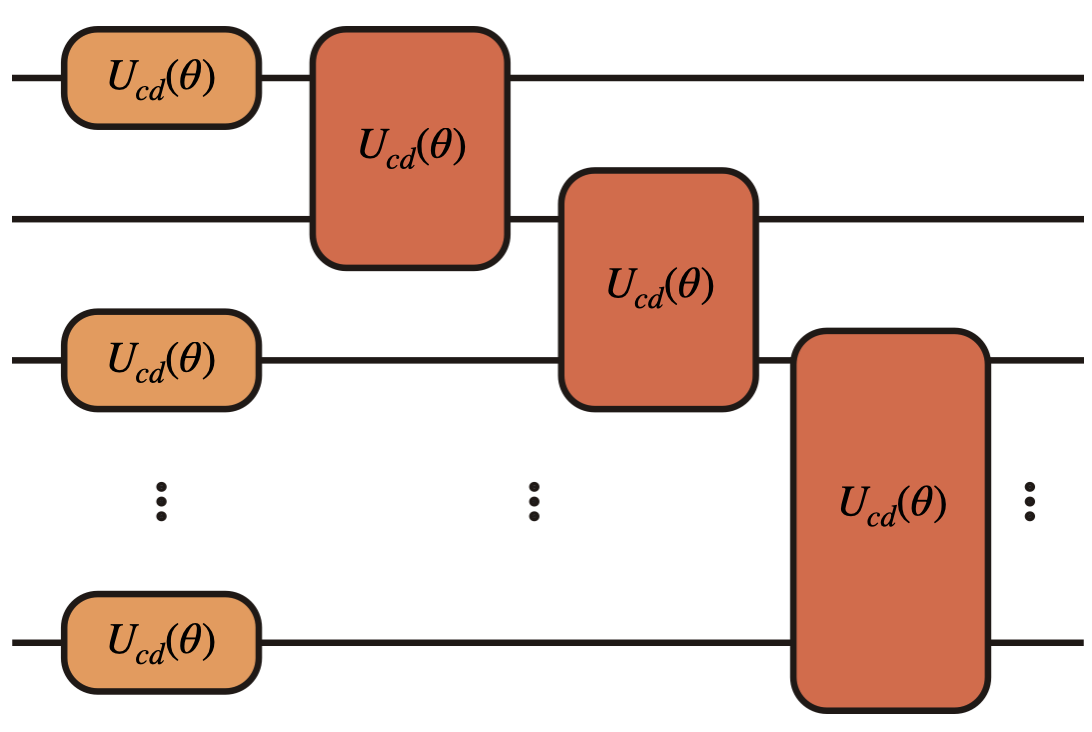}
        \caption{CD-QAOA circuit for a certain layer.}
        \label{fig:cd_qaoa_circuit}
    \end{subfigure}

    \caption{Schematic representation of the tetrahedral lattice and CD-QAOA circuit.}
    \label{fig:cdqaoa}
\end{figure}

\subsubsection{Quantum Approximate Optimization Algorithms (QAOAs)}

We define the Hamiltonian as $H_{gc}+H_{int}$. One way to minimize the Hamiltonian is QAOA. It is inspired by adiabatic computing, which uses a combination of the problem Hamiltonian $H_p$ and the initial mixing Hamiltonian $H_m$.
\begin{equation}
H_a(t) = (1-\lambda(t))H_{m} + \lambda(t) H_p.
\end{equation}

Recently, extensive efforts have been devoted to accelerating convergence using counter-diabatic (CD) QAOAs, which introduce a Hamiltonian including a CD term as follows:
\begin{equation}
H_{cd}(t) = (1-\lambda(t))H_{m} + \lambda(t) H_p + \dot{\lambda}A^l_\lambda 
\end{equation}

where $A^l_\lambda$ is called an adiabatic gauge potential.
\begin{equation}
A^l_\lambda = i \sum_{k=1}^l \alpha_k(t) 
\underbrace{[H_a, [H_a, \cdots [H_a,\partial_\lambda H_a ]]]}_{2k-1 \text{ commutators}}
\label{Al}
\end{equation}

Note that we have used $[A,B]=AB-BA$ as a commutator. As $l$ increases, the term becomes closer to an ideal CD term. Due to the complexity of the $l \to \infty$ case, we perform calculations only up to $l=2$. Following the procedure \cite{Chandarana2022}, when $\dot{\lambda}$ is sufficiently large, $H_{cd}$ can be approximated using only the counter-diabatic terms.\\

\subsubsection{Implementing the quantum circuit}

We calculated $H_{gc}+H_{int}$, which is represented as a polynomial of $q_0,q_1,\cdots$.\\
\begin{equation}
H_{gc}+H_{int}=\sum_a G_a q_a + \sum_{a<b} G_{ab} q_a q_b
+ \cdots + \sum_{a<b<c<d<e} G_{abcde} q_a q_b q_c q_d q_e
\end{equation}

We convert this polynomial into a problem Hamiltonian that contains the Pauli $Z$ basis. Technically, $q_i$ maps to $Z_i$, though the coefficients may vary. There are up to five-body $Z$ interactions, such as $Z_a Z_b Z_c Z_d Z_e$.\\
\begin{equation}
H_p=\sum_a J_a Z_a + \sum_{a<b} J_{ab}Z_a Z_b
+ \cdots + \sum_{a<b<c<d<e} J_{abcde}Z_a Z_b Z_c Z_d Z_e
\end{equation}

Since $H_p$ only includes $Z_i$ terms, its components commute. We can build a unitary operator with parameters $\gamma_l$ $(1\leq l \leq p)$.\\
\begin{equation}
e^{-i \gamma_l H_p}=\prod_a J_a e^{-i \gamma_l Z_a}  \prod_{a<b} J_{ab} e^{-i\gamma_l Z_a Z_b}
\cdots  \prod_{a<b<c<d<e} J_{abcde} e^{-i \gamma_l Z_a Z_b Z_c Z_d Z_e}
\label{UHp}
\end{equation}

The mixing Hamiltonian $H_m$ is prepared using the Pauli $X$ basis.\\
\begin{equation}
H_m= \sum_a X_a
\end{equation} 

Likewise, its unitary form is\\
\begin{equation}
e^{-i \beta_l H_m} =\prod_a e^{-i \beta_l X_a} 
\label{UHm}
\end{equation} 

which is realized by single-qubit $X$-rotation gates. Therefore, $H_a = (1-\lambda(t))H_{m} + \lambda(t)H_p$ contains Pauli $X_a$ and $Z_b$ terms, for example, $X_1, X_2, \cdots$ and $Z_1,Z_2,Z_3,\cdots,Z_1Z_2Z_3Z_4Z_5,\cdots$. By calculating Eq.~(\ref{Al}), we obtain many Pauli matrix combinations.\\

Considering the complexity, we performed calculations only up to $l=2$. Following \cite{Chandarana2022}, we chose $Y$ and $YZ$ terms, which are converted into a single-qubit $Y$-rotation gate and a $YZ$-rotation gate on the quantum circuit. They made an approximation that, for a sufficiently large $\dot{\lambda}$ during CD evolution, the adiabatic terms can be ignored. The unitary form is\\
\begin{equation}
e^{-i \theta_l A^l_{\lambda}} =\prod_a K_a e^{-i \theta_a Y_a}  \prod_{a<b} K_{ab} e^{-i \theta_{ab} Y_a Z_b}
\label{Ucd}
\end{equation} 

The problem Hamiltonian and mixing Hamiltonian include only one type of Pauli matrix, and their components all commute. Therefore, in the same layer, the unitary gate is realized using only one parameter, as shown in Eqs.~(\ref{UHp}) and (\ref{UHm}). However, for CD-QAOA, $A^l_{\lambda}$ contains both Pauli $X$ and Pauli $Z$, and some of its terms do not commute. Therefore, CD-QAOA unitary gates cannot be represented by a single parameter in a layer. Instead, we assign independent parameters $\Theta_i$, where\\
\begin{equation}
\theta_a \in \mathbf{\Theta}_i \quad \text{for each } a, 
\quad 
\theta_{ab} \in \mathbf{\Theta}_i \quad \text{for each } a,b.
\end{equation}

Here, $i$ denotes the $p=i$ layer. The CD-QAOA quantum circuit evolves the quantum state as\\
\begin{equation}
|\psi_f \rangle = U_{cd}(\mathbf{\Theta}_p) U_b(\beta_p) U_c(\gamma_p) \cdots U_{cd}(\mathbf{\Theta}_1) U_b(\beta_1) U_c(\gamma_1) |\psi_i \rangle 
\end{equation}

Ignoring the adiabatic terms, it becomes\\
\begin{equation}
|\psi_f \rangle = U_{cd}(\mathbf{\Theta}_p)  \cdots U_{cd}(\mathbf{\Theta}_1) |\psi_i \rangle.
\end{equation}

Fig.~\ref{fig:cd_qaoa_circuit} describes the QAOA quantum circuit. The unitary gates (colored in  orange) represent single-qubit $Y$-rotation gates, while the unitary gates (colored in red) represent $YZ$ interactions acting on two qubits.\\

\subsection{ Hartree-Fock simulation  }
The Hartree--Fock (HF) method is based on the mean-field approximation, in which each electron is assumed to move in the average electrostatic field generated by all other electrons \cite{Szabo1996}. It is particularly effective for determining the fundamental electronic configuration of a system and provides a reliable reference wavefunction for more advanced methods, such as post-HF approaches and DFT.\\

In HF theory, the electrons are described by an antisymmetrized product, known as a Slater determinant. Each electron moves independently of all the others under the Coulomb repulsion arising from the average positions of all electrons, together with exchange interactions. The one-electron operator $h$ is given as
\begin{equation}
h(i) = -\frac{1}{2}\nabla_i^2 - \sum_A \frac{Z_A}{r_{iA}}
\end{equation}
Here, $Z_A$ is the nuclear charge of nucleus $A$, and $r_{iA}$ is the distance between electron $i$ and nucleus $A$.
The two-electron operator $v(i,j)$ is given as
\begin{equation}
v(i,j) = \frac{1}{r_{ij}}
\end{equation}
The electronic energy in the HF method is given as
\begin{equation}
E_{HF} = \sum_i \langle i | h | i \rangle 
+ \frac{1}{2} \sum_{ij} \left( \langle ij | ij \rangle - \langle ij | ji \rangle \right)
\end{equation}
where the one-electron and two-electron integrals are given as
\begin{equation}
\langle i|h|j\rangle = \int d\mathbf{x}_1 \chi_i^*(\mathbf{x}_1)h(\mathbf{r}_1)\chi_j(\mathbf{x}_1)
\end{equation}
\begin{equation}
\langle ij|kl \rangle= \int d\mathbf{x}_1 d\mathbf{x}_2 \chi_i^*(\mathbf{x}_1)\chi_j(\mathbf{x}_1)\frac{1}{r_{12}}\chi_k^*(\mathbf{x}_2)\chi_l(\mathbf{x}_2)
\end{equation}
where $\chi_i$ is a spin orbital. Compared with DFT or post-HF methods, such as MP2 (Second-order M{\o}ller--Plesset perturbation theory \cite{Moller1934}) or CCSD(T) (Coupled cluster with single, double, and perturbative triple excitations \cite{Raghavachari1989, Stanton1997}), HF is computationally less demanding. Moreover, because HF treats the exchange energy exactly at the mathematical level, it does not suffer from the self-interaction error (SIE), a well-known limitation of DFT \cite{Perdew1981}. SIE arises when an electron is spuriously allowed to interact with itself, leading to an artificial increase in energy when electrons are localized. As a consequence, DFT often exhibits over-delocalization, where electrons are artificially spread out more than they should be in order to minimize this error. This can result in an inaccurate description of charge-separated states. Such inaccuracies can distort key interactions in peptides, including hydrogen bonding and charge transfer, which are critical for determining secondary structure. In systems where charge separation plays an important role, HF can yield results that are more physically meaningful than those obtained from DFT.\\

Therefore, for relatively short peptides containing more than seven amino acids, such as APRLRFY considered in this study, the number of atoms is still sufficiently large to make high-level calculations expensive. In this work, HF/6-31G was employed to rapidly capture the overall structural trends through coarse-grained optimization.

\subsection{Density Functional Theory (DFT) Calculation}

Density Functional Theory (DFT) is a widely used quantum chemical method for calculating the electronic structure and energy of molecular systems. Unlike the Hartree--Fock (HF) method, which describes an electronic system using the many-electron wavefunction $\Psi$, DFT uses the electron density $\rho(\mathbf{r})$ as the fundamental variable \cite{Hohenberg1964}. The electron density represents the spatial distribution of electrons and can be expressed using the Kohn--Sham orbitals as \cite{Kohn1965}
\begin{equation}
\rho(\mathbf{r})
=
\sum_i^{\mathrm{occ}}
n_i
\left|
\psi_i(\mathbf{r})
\right|^2 ,
\end{equation}
where $\psi_i(\mathbf{r})$ is the $i$th Kohn--Sham orbital and $n_i$ is its occupation number. In Kohn--Sham DFT, the total energy is expressed as a functional of the electron density:
\begin{equation}
E_{\mathrm{KS}}[\rho]
=
T_s[\rho]
+
E_{\mathrm{ne}}[\rho]
+
J[\rho]
+
E_{\mathrm{xc}}[\rho] ,
\end{equation}
where $T_s[\rho]$ is the kinetic energy of the non-interacting reference system, $E_{\mathrm{ne}}[\rho]$ is the nucleus--electron attraction energy, $J[\rho]$ is the classical electron--electron Coulomb repulsion energy, and $E_{\mathrm{xc}}[\rho]$ is the exchange--correlation energy. The exchange--correlation term accounts for the non-classical exchange and correlation effects between electrons, and the accuracy of DFT calculations strongly depends on the choice of the exchange--correlation functional.\\

The Kohn--Sham orbitals are obtained by solving the Kohn--Sham equations self-consistently:
\begin{equation}
\left[
-\frac{1}{2}\nabla^2
+
v_{\mathrm{eff}}(\mathbf{r})
\right]
\psi_i(\mathbf{r})
=
\varepsilon_i
\psi_i(\mathbf{r}) ,
\end{equation}
where $v_{\mathrm{eff}}(\mathbf{r})$ is the effective potential, including the nucleus--electron attraction, classical Coulomb repulsion, and exchange--correlation potential.\\

Compared with HF, DFT can include electron correlation effects through the exchange--correlation functional, making it practical for geometry optimization and electronic structure analysis of organic molecules and peptide systems. In addition, compared with force-field-based molecular dynamics approaches, DFT is less dependent on empirical parameterization and can explicitly describe electronic structure effects such as bond polarization and charge redistribution between residues. Therefore, DFT provides a useful framework for evaluating potential energy surfaces and non-covalent interactions in peptide systems.\\

Short peptides, typically composed of approximately 5--20 amino acid residues, often exhibit highly flexible conformational landscapes. Unlike large proteins that adopt relatively stable and well-defined tertiary structures, short peptides may have multiple accessible conformations, and their relative stability is governed by a delicate balance of electrostatic interactions, hydrogen bonding, backbone conformational preferences, side-chain packing, and other non-covalent inter-residue interactions. For this reason, quantum chemical calculations can be useful for investigating the ground-state conformations of short peptides.\\

However, conventional DFT methods have limitations in accurately describing long-range dispersion, or van der Waals, interactions. This issue is particularly important for peptide systems, where weak non-covalent interactions contribute significantly to backbone folding, side-chain packing, and hydrophobic interactions. To improve the description of these interactions, dispersion-corrected DFT methods such as DFT-D3, DFT-D4, and empirical-dispersion-containing functionals have been developed \cite{grimme2010,Caldeweyher2019DFTD4,Chai2008WB97XD}. In this study, the B3LYP functional with Grimme's D3 dispersion correction and Becke--Johnson damping was used to account for dispersion interactions in the APRLRFY peptide \cite{grimme2010,grimme2011}. The total energy was expressed as
\begin{equation}
E_{\mathrm{B3LYP-D3BJ}}
=
E_{\mathrm{B3LYP}}
+
E_{\mathrm{disp}}^{\mathrm{D3BJ}} .
\end{equation}
Geometry optimization was performed by minimizing the electronic energy with respect to the nuclear coordinates $\mathbf{R}$:
\begin{equation}
\mathbf{R}_{\mathrm{opt}}
=
\operatorname*{arg\,min}_{\mathbf{R}}
E(\mathbf{R}) .
\end{equation}
DFT calculations were performed to predict the optimized structure of the APRLRFY peptide. All calculations were carried out using Gaussian 16 \cite{Frisch2016}. Frequency calculations were additionally performed to confirm that the optimized structure corresponds to a local minimum on the potential energy surface, as indicated by the absence of imaginary frequencies.

\subsection{ MD simulation }
Molecular dynamics (MD) simulation is a classical computational method used to investigate the time-dependent motion of atoms and molecules \cite{AllenTildesley2017}. In contrast to quantum chemical methods such as DFT, which calculate molecular structures based on electronic structure, MD simulations describe molecular motion using Newton's equations of motion:
\begin{equation}
m_i \frac{d^2 \mathbf{r}_i}{dt^2}
=
\mathbf{F}_i
\end{equation}
The force acting on each atom is obtained from the gradient of the potential energy function:
\begin{equation}
\mathbf{F}_i
=
-
\nabla_i U(\mathbf{R})
\end{equation}
In MD simulations, the potential energy $U(\mathbf{R})$ is described using a molecular mechanics force field, which consists of bonded and nonbonded interaction terms \cite{Ponder2003,MacKerell1998}:
\begin{equation}
U_{\mathrm{total}}
=
U_{\mathrm{bonded}}
+
U_{\mathrm{nonbonded}}
\end{equation}
A general force field can be written as
\begin{equation}
\begin{aligned}
U_{\mathrm{total}}
=&
\sum_{\mathrm{bonds}}
k_b (r-r_0)^2
+
\sum_{\mathrm{angles}}
k_\theta (\theta-\theta_0)^2
+
\sum_{\mathrm{dihedrals}}
k_\phi
\left[
1+\cos(n\phi-\delta)
\right] \\
&+
\sum_{i<j}
\left[
4\varepsilon_{ij}
\left\{
\left(
\frac{\sigma_{ij}}{r_{ij}}
\right)^{12}
-
\left(
\frac{\sigma_{ij}}{r_{ij}}
\right)^6
\right\}
+
\frac{q_i q_j}{4\pi \varepsilon_0 r_{ij}}
\right].
\end{aligned}
\end{equation}
The bonded terms describe bond stretching, angle bending, and dihedral rotation, whereas the nonbonded terms describe van der Waals and electrostatic interactions \cite{Ponder2003,Vanommeslaeghe2014}. By numerically integrating the equations of motion, MD simulations generate atomic trajectories that can be used to analyze the conformational flexibility and dynamic structural behavior of peptides and proteins.

\subsection{Hamiltonian replica exchange molecular dynamics (H-REMD)}
Replica exchange molecular dynamics (REMD) is a method that avoids trapping in local minimum states by employing a generalized-ensemble algorithm, which uses non-Boltzmann probability weight factors to weight each state. During the REMD simulation, non-interacting replicas at different temperatures exchange configurations according to a specified transition probability. The transition probability of the replica-exchange process between two different states, $X$ and $X'$, is given by the following equation \cite{han2014protein}:\\
\begin{equation}
    w(X, X') = \min \left( 1, \exp(-\Delta) \right)
    \label{eq:transition_probability}
\end{equation} 
\begin{equation}
    \Delta = \beta_m \left( E(q^{[i]}) - E(q^{[j]}) \right) - \beta_n \left( E(q^{[j]}) - E(q^{[i]}) \right) = (\beta_m - \beta_n) \left( E(q^{[j]}) - E(q^{[i]}) \right)
    \label{eq:delta_equation}
\end{equation}
where $\beta_m = {1}/{k_B T_m}$, $\beta_n = {1}/{k_B T_n}$, $k_B$ is the Boltzmann constant, $q^{[i]}$ is the coordinate of replica $i$, and $q^{[j]}$ is the coordinate of replica $j$.\\

Here, the potential energy $E(q^{[i]})$ is composed of intraprotein interaction energy, protein--solvent interaction energy, and water--water interaction energy of replica $i$. This temperature-dependent REMD method is effective for identifying the lowest-energy conformation of a small peptide solvated in water, but a considerably large number of replicas is required to efficiently sample the conformational space, which scales as $O(f^{1/2})$, where $f$ is the number of degrees of freedom in the system \cite{fukunishi2002hamiltonian}.\\

Replica exchange with solute tempering (REST) was devised as a method that considers only the solute system when calculating replica-exchange acceptance probabilities \cite{wang2011replica}.\\
\begin{equation}
    \Delta = (\beta_m - \beta_n) \left(E(q^{[j]}) - E(q^{[i]})\right) = 
     (\beta_m - \beta_n) \left[(E_{pp}(q^{[j]}) + \frac{1}{2}E_{pw}(q^{[j]})) - (E_{pp}(q^{[i]}) + \frac{1}{2}E_{pw}(q^{[i]}))\right]
    \label{eq:delta_equation_rest}
\end{equation}
where $E_{pp}$ and $E_{pw}$ are the potential energies of intraprotein interactions and protein--solvent interactions, respectively.\\

This significantly increases the efficiency of REMD calculations by removing the water--water self-interaction energy from the acceptance criterion, which is the most dominant term among the three interaction terms due to the large number of water molecules in the system \cite{huang2007replica}. Thus, the conformational space that must be searched is reduced from $O(f^{1/2})$ to $O(f_p^{1/2})$, where $f_p$ is the number of degrees of freedom of the main solute system, namely, the protein. However, the authors who developed the REST method found that systems undergoing large conformational changes, such as folded-to-unfolded transitions of proteins, did not show increased efficiency compared with the original temperature REMD method \cite{huang2007replica}. This has been ascribed to the fact that water molecules near the protein compensate for protein--water interactions and promote the exchange of configurations at different temperature levels \cite{huang2007replica}. In REST1, the potential energy at temperature $m$ and coordinate $q^{[i]}$ is determined as follows \cite{wang2011replica}:
\begin{equation}
    E_m^{(q^{[i]})} = E_{pp}(q^{[i]}) + \frac{\beta_0 + \beta_m}{2\beta_m} E_{pw}(q^{[i]}) + \frac{\beta_0}{\beta_m} E_{ww}(q^{[i]})
    \label{eq:placeholder}
\end{equation}
In REST2, which is Hamiltonian replica exchange molecular dynamics, the intraprotein interaction energy is scaled by $\beta_m/\beta_0$. The barrier separating two different conformations is lowered by this scaling of the energy term. In REST1, this barrier is not changed, and only the temperature is increased to overcome the energy barrier. On the other hand, the scaling factor for the protein--water interaction energy term is $\sqrt{{\beta_m}/{\beta_0}}$, which is different from ${\left( \beta_0 + \beta_n \right)}/{2 \beta_m}$ in REST1. This adjusted scaling of the protein--water interaction energy term also affects the performance of REST2. The acceptance probability of REST2 is calculated by the following equation \cite{wang2011replica}:
\begin{equation}
    \Delta = (\beta_m - \beta_n) \left[ \left( E_{pp}(q^{[j]}) + \frac{\sqrt{\beta_0}}{\sqrt{\beta_m} + \sqrt{\beta_n}} E_{pw}(q^{[j]}) \right) - \left( E_{pp}(q^{[i]}) + \frac{\sqrt{\beta_0}}{\sqrt{\beta_m} + \sqrt{\beta_n}} E_{pw}(q^{[i]}) \right) \right]
    \label{eq:complex_delta_formula}
\end{equation}
As seen from the equation, the scaling term of REST2 was changed from $1/2$ in REST1 to ${\sqrt{\beta_0}}/({\sqrt{\beta_m} + \sqrt{\beta_n}})$. In REST2, the potential energy of coordinate $q^{[i]}$ at temperature $m$ follows the equation below:
\begin{equation}
    E_m^{(q^{[i]})} = \frac{\beta_m}{\beta_0} E_{pp}(q^{[i]}) + \sqrt{\frac{\beta_m}{\beta_0}} E_{pw}(q^{[i]}) + E_{ww}(q^{[i]})
    \label{eq:complex_energy_equation}
\end{equation}
We generated $4$ scaling values and applied them to the solvent molecules by rescaling the force field. Coordinate exchange was attempted every $400$ steps, and each replica was run for $500$ ns; therefore, a total of $2~\mu$s of simulations were conducted.\\
\section{Method}
\subsection{A CD–QAOA/MD Hybrid Method}
To describe residue--residue interactions more specifically, the Miyazawa--Jernigan (MJ) interaction matrix \cite{Miyazawa1985} was incorporated into the Hamiltonian of the initial CD-QAOA model. In the first CD-QAOA model, the interaction coefficient between residues 2 (Pro) and 7 (Tyr) was treated as a fixed value.
In the 2nd CD-QAOA model, the pairwise interaction energy term was instead calculated using amino-acid-pair-specific MJ energy values. By introducing the MJ potential, residue-specific interactions, including hydrophobic interactions, polar residue interactions, and charged residue interactions, were explicitly reflected in the Hamiltonian. This modification allowed the model to represent the physicochemical characteristics of residue--residue contacts more realistically during the folding process.\\

Following the introduction of the MJ interaction term, the length of the final bitstring was extended from 12 bits to 14 bits. The initial 12 bits mainly represented backbone movement directions or lattice-based structural states of the protein, whereas the two additional bits were used to represent information associated with the interaction term.\\

The continuity penalty parameter $\lambda_{2}$ was adjusted because the inclusion of residue-specific attractive interactions, such as those represented by the MJ potential, may cause strong stabilizing interactions between specific residue pairs to dominate over the structural constraint terms. Such dominance can lead to nonphysical local minima, including chain discontinuity, unrealistic lattice movements, or structural crossings. Therefore, a sufficiently large structural penalty was introduced to suppress physically invalid conformations and maintain the continuity of the protein backbone during the optimization process.\\

The bitstring obtained from CD-QAOA was first converted into a coarse-grained APRLRFY peptide structure represented by $C_{\alpha}$ atoms. Since the initial bitstring-derived structure contained only $C_{\alpha}$ coordinates, reconstruction of an all-atom structure was required for further structural analysis and molecular dynamics simulations. Therefore, PULCHRA was used to rebuild the backbone and side-chain atoms from the $C_{\alpha}$-only model \cite{rotkiewicz2008pulchra}. The reconstructed all-atom APRLRFY peptide structure, including side chains, was then used as the initial structure for the subsequent 10 ns MD simulation in explicit water environment and stabilized through energy minimization and equilibration processes.\\

In the MD simulations, interatomic interactions were calculated using the bond, angle, dihedral angle, and nonbonded interaction terms included in the force field. In particular, the dihedral angle potential controls the rotational state of the peptide backbone and side-chain orientation, thereby suppressing unrealistic structural deformation. Additionally, the stereochemical parameters of the force field, including improper dihedral terms, contribute to maintaining the stereochemical arrangement of chiral centers; thus, the chirality of the peptide can be reflected during the MD stabilization process \cite{MacKerell1998, Hills2014}.\\

Therefore, in this study, by adding an MD stabilization process to the CD-QAOA-based candidate structures, final candidate conformations were derived that consider the stable structure of the side chains, in the aqueous environment, and force-field-based stereochemical conditions.

\subsection{Hartree Fock calculation}
Hartree--Fock (HF) calculations for the heptapeptide sequence APRLRFY were performed using Gaussian 16 \cite{Frisch2016}. The initial peptide structure was constructed in a fully extended conformation using UCSF ChimeraX \cite{Pettersen2021}, with all backbone dihedral angles set to $\phi = 180^\circ$ and $\psi = 180^\circ$. This configuration minimizes initial steric hindrance and enables a clearer assessment of structural changes driven by electrostatic repulsion between positively charged residues. The unrestricted Hartree--Fock (UHF) formalism was employed for electronic structure optimization and self-consistent field (SCF) calculations, allowing independent optimization of the $\alpha$ and $\beta$ spin orbitals. All calculations were carried out in the gas phase. The total charge of the system was set to $+2$ to account for the protonated states of the two arginine residues, and the spin multiplicity was specified as 1 (singlet). All atoms were described using the 6-31G split-valence basis set, which offers a practical balance between computational efficiency and accuracy and is widely adopted in quantum chemical studies.

\subsection{DFT calculation}
DFT-based structural optimization and frequency calculations were performed to predict the structure of the APRLRFY peptide. All calculations were conducted using the Gaussian 16 program with the B3LYP exchange--correlation functional and the 6-31G(d) basis set. The B3LYP/6-31G(d) level of theory was selected to balance computational accuracy and efficiency in peptide structure optimization. Geometry optimization and frequency calculations were performed sequentially at the same computational level. The frequency calculations were conducted to evaluate the vibrational characteristics of the optimized structure and to assess whether the optimized geometry corresponded to an energy minimum. The aqueous environment was modeled using the SMD (Solvation Model based on Density) implicit solvation model, with water selected as the solvent \cite{marenich2009}. Additionally, Grimme's D3 dispersion correction with Becke--Johnson damping (GD3BJ) was applied to account for long-range dispersion interactions \cite{grimme2010,grimme2011}. During the calculations, the total charge and spin multiplicity of the APRLRFY peptide were set to $+2$ and $1$, respectively.

\subsection{Conventional MD simulation for energy minimization and relaxation}

The initial structure of the APRLRFY peptide was predicted using AlphaFold3 \cite{Abramson2024}. Steric clashes were identified through structural inspection. These clashes were resolved by energy minimization followed by molecular dynamics (MD) simulations, resulting in a structurally stable conformation. MD simulations were performed using the NAMD 2.14 package \cite{Phillips2020} with the CHARMM36 force field parameters \cite{Best2012}. The system was solvated in a TIP3P water box \cite{Jorgensen1983} ($42.37~\text{\AA} \times 42.51~\text{\AA} \times 46.63~\text{\AA}$) , and 0.15 M NaCl was added for neutralization. A cutoff distance of $12~\text{\AA}$ was applied for nonbonded interactions, and long-range electrostatic interactions were treated using the particle mesh Ewald (PME) method with a grid spacing of $1~\text{\AA}$ \cite{darden1993particle}. The system pressure was controlled using the Langevin dynamics method \cite{Schneider1978}. Following energy minimization and heating, equilibration was carried out in the NPT ensemble at 310 K and 1 atm. Initial restraints were gradually released to allow full structural relaxation and stabilization of the system. A 10 ns MD simulation was performed to assess the structural stability of the peptide.

\subsection{MD simulation for H-REMD}

Molecular dynamics (MD) simulations were performed using GROMACS with two AMBER force fields, ff99SB-ILDN \cite{lindorff2010improved} and ff99SB-DISP \cite{robustelli2018developing}. The ff99SB-ILDN simulations employed the TIP3P water model, whereas ff99SB-DISP was used in conjunction with the modified TIP4P water model \cite{Jorgensen1983}. All bonds involving hydrogen atoms were constrained using the LINCS algorithm \cite{hess1997lincs}, allowing the use of a 2 fs integration time step. The equations of motion were integrated using the Verlet algorithm.\\

A cutoff of 10 $\text{\AA}$ was applied for short-range electrostatic and van der Waals interactions. Long-range electrostatics were treated using the particle mesh Ewald (PME) method with fourth-order cubic interpolation and a Fourier grid spacing of $1.6\text{\AA}$ \cite{darden1993particle}. The system was solvated in a cubic box with a side length of $55.4 \text{\AA}$, resulting in a total of 21,816 atoms. Sodium ($\mathrm{Na}^{+}$) and chloride ($\mathrm{Cl}^{-}$) ions were added to ensure overall charge neutrality.\\

Energy minimization was carried out using the steepest descent algorithm with a maximum of 50,000 steps \cite{Nocedal2006}, without applying constraints. The system was subsequently equilibrated in the NVT ensemble for 0.1 ns using the V-rescale thermostat at a reference temperature of 300 K. This was followed by 0.1 ns equilibration in the NPT ensemble using the Parrinello--Rahman barostat at a reference pressure of 1 bar \cite{Parrinello1981}. Finally, a 1 ns production run was conducted in the NPT ensemble. All simulations were performed with a time step of 2 fs.

\vspace{1em}
\section{Results and Discussion}
\subsection{Prediction of a protein structure for $N=7$ using QAOAs}

\begin{figure}[h]
    \centering

    \begin{subfigure}{0.45\textwidth}
        \centering
        \includegraphics[width=\linewidth]{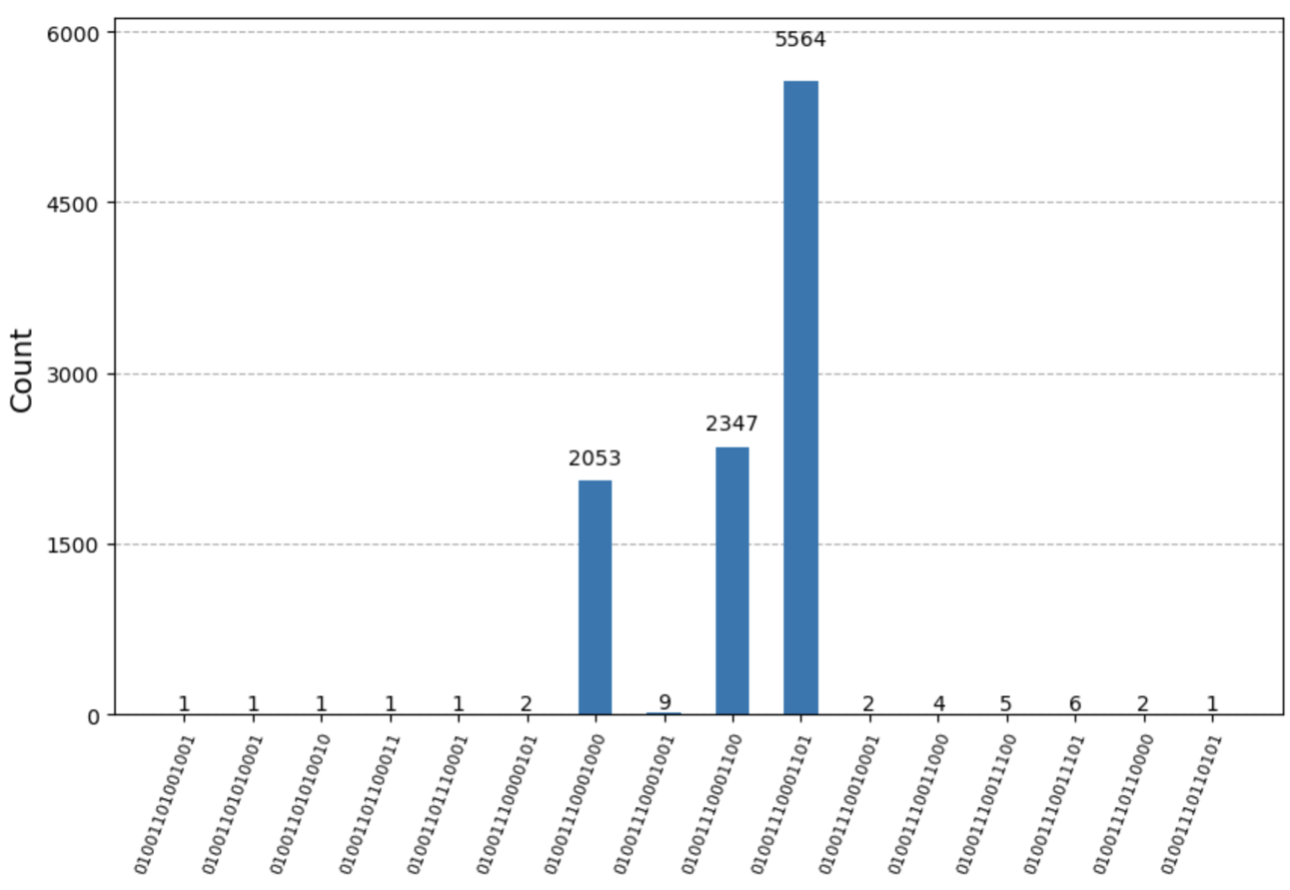}
        \caption{Probability distribution of CD-QAOA at $l=1$.}
        \label{fig:cdqaoa_l1_pd}
    \end{subfigure}
    \hfill
    \begin{subfigure}{0.45\textwidth}
        \centering
        \includegraphics[width=\linewidth]{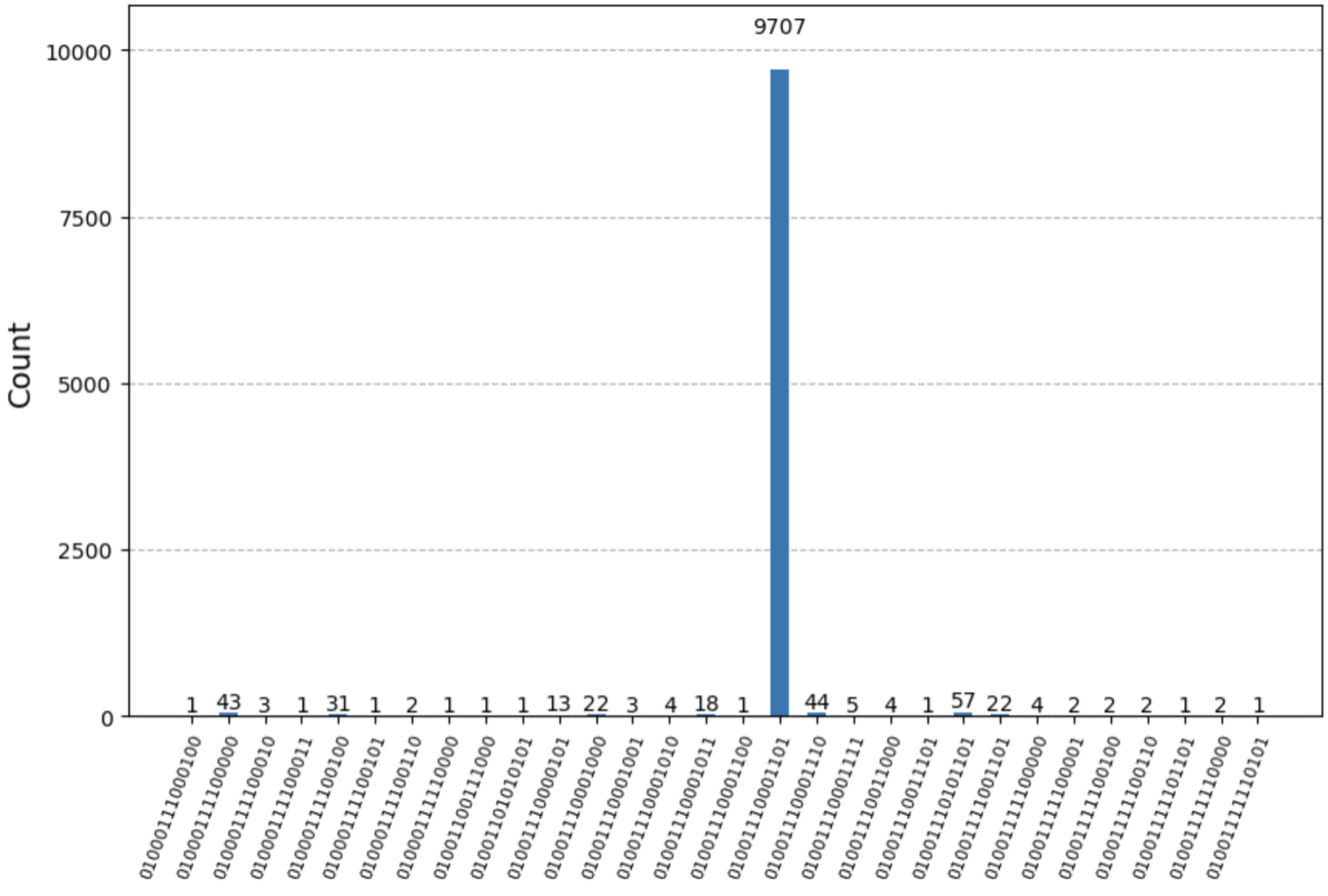}
        \caption{Probability distribution of CD-QAOA at $l=2$.}
        \label{fig:cdqaoa_l2_pd}
    \end{subfigure}

    \vspace{0.15cm}

    \begin{subfigure}{0.45\textwidth}
        \centering
        \includegraphics[width=\linewidth]{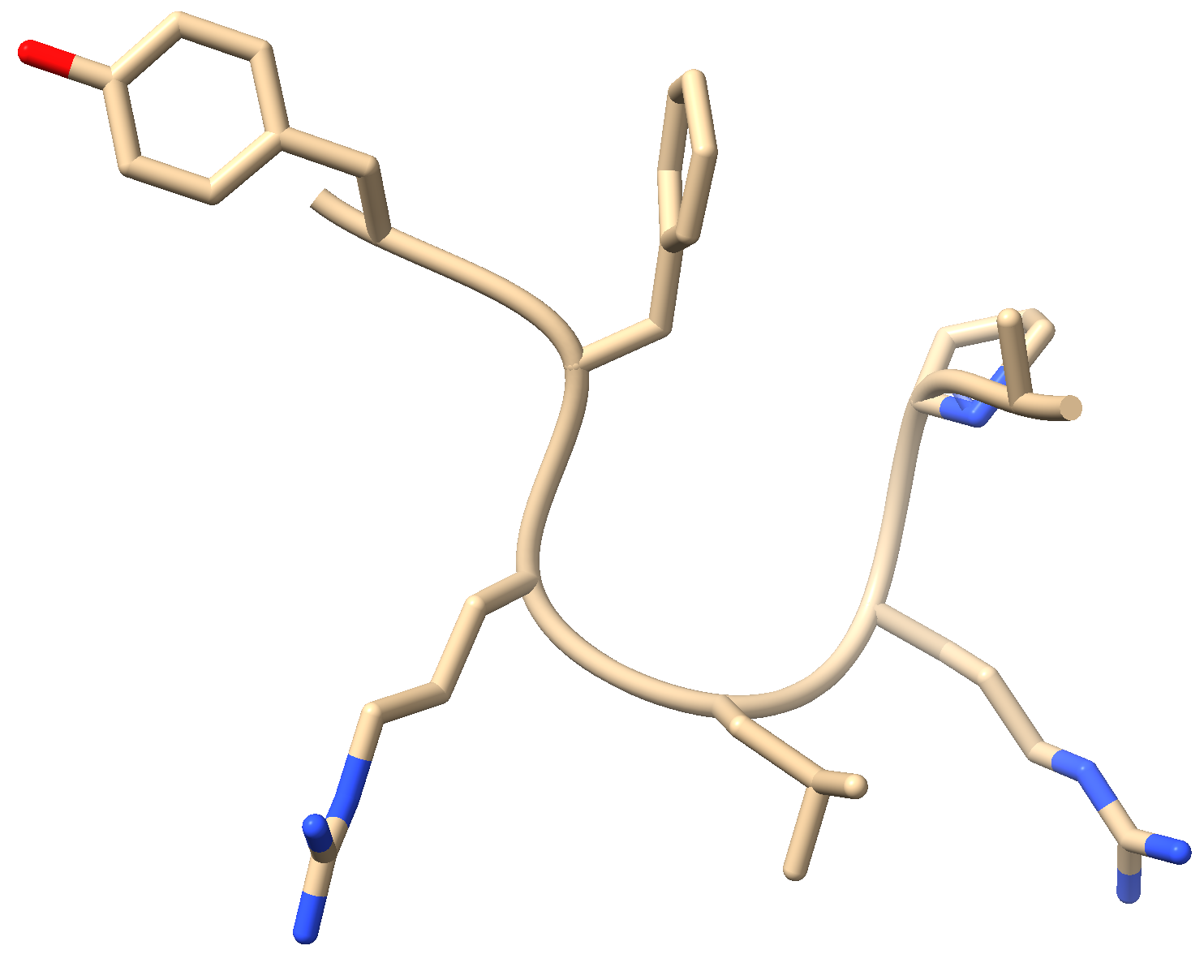}
        \caption{APRLRFY $C_{\alpha}$ structure generated from the bitstring 010011100010.}
        \label{fig:apr_010011100010}
    \end{subfigure}
    \hfill
    \begin{subfigure}{0.45\textwidth}
        \centering
        \includegraphics[width=\linewidth]{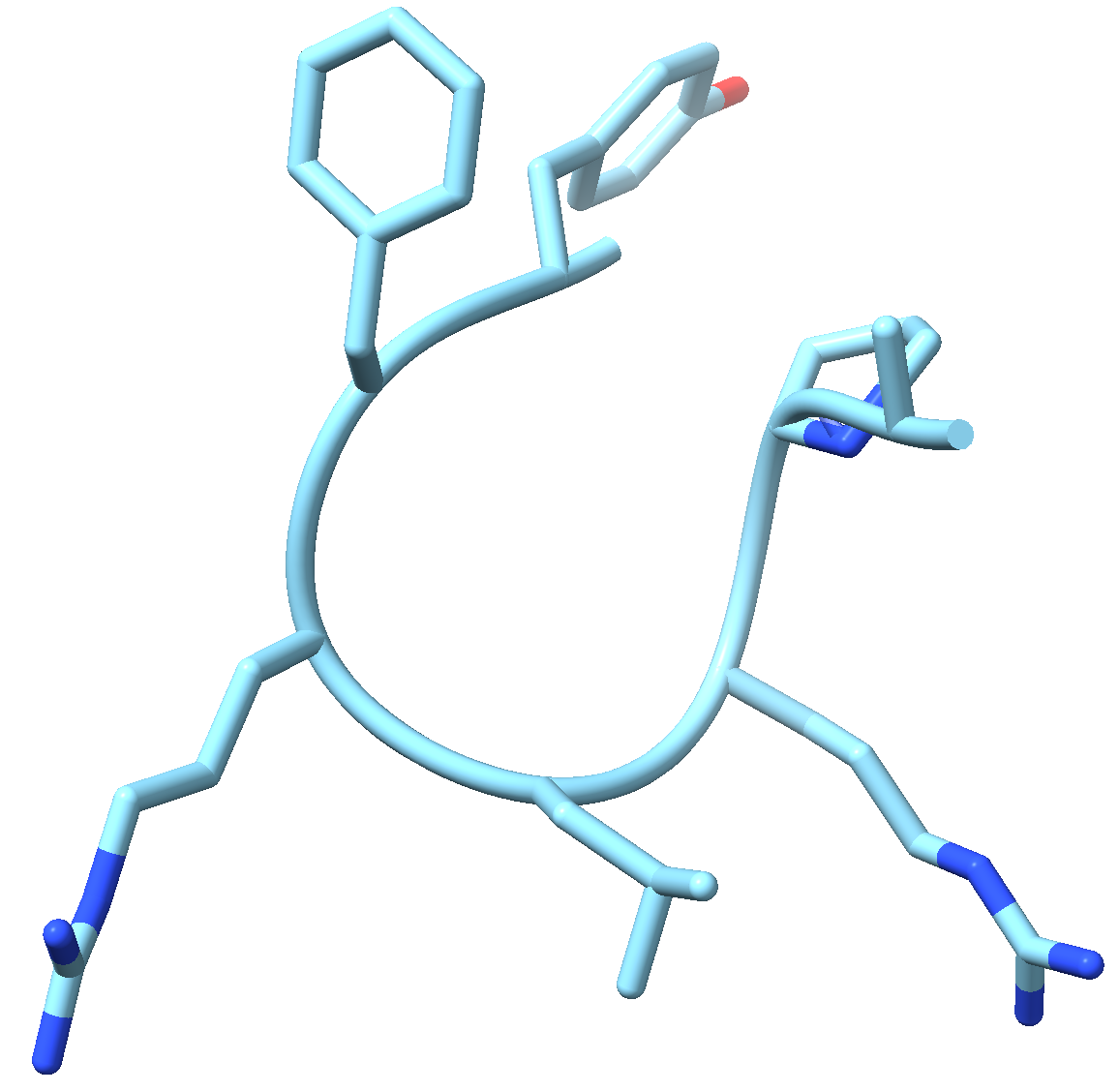}
        \caption{APRLRFY $C_{\alpha}$ structure generated from the bitstring 010011100011.}
        \label{fig:apr_010011100011}
    \end{subfigure}

    \caption{
    Probability distributions of CD-QAOA and reconstructed APRLRFY $C_{\alpha}$ structures.
    The optimal bitstring 010011100010 was obtained from the CD-QAOA case at $l=1$, whereas 010011100011 was obtained from both the $l=1$ and $l=2$ cases.
    The $C_{\alpha}$ structures were reconstructed using the method \cite{rotkiewicz2008pulchra}.
    }
    \label{fig:cdqaoa_aprlrfy_result}
\end{figure}

\begin{figure}[t]
    \centering

    \begin{subfigure}{0.48\textwidth}
        \centering
        \includegraphics[width=\linewidth]{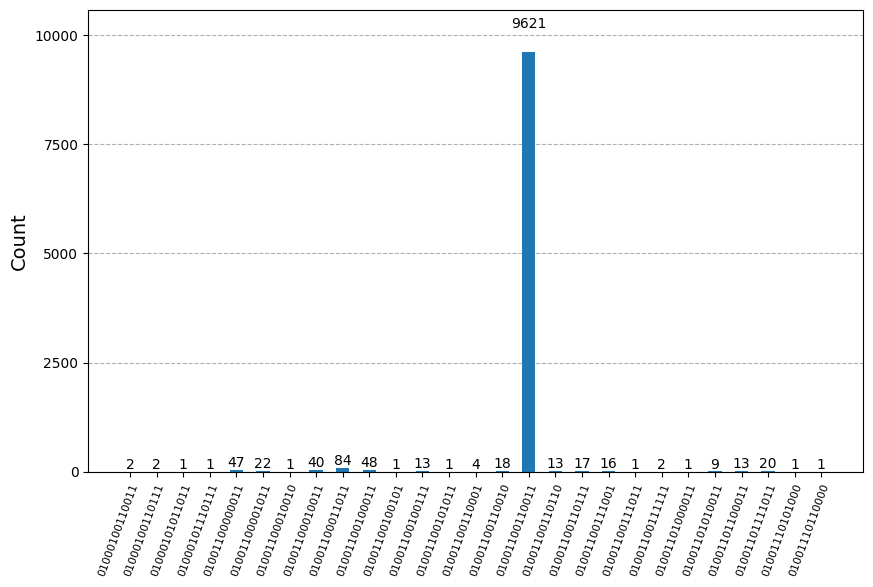}
        \caption{Probability distribution obtained from the CD-QAOA simulation with the MJ interaction at $l=1$.}
        \label{fig:cdqaoa_l1_pd_mj}
    \end{subfigure}
    \hfill
    \begin{subfigure}{0.48\textwidth}
        \centering
        \includegraphics[width=\linewidth]{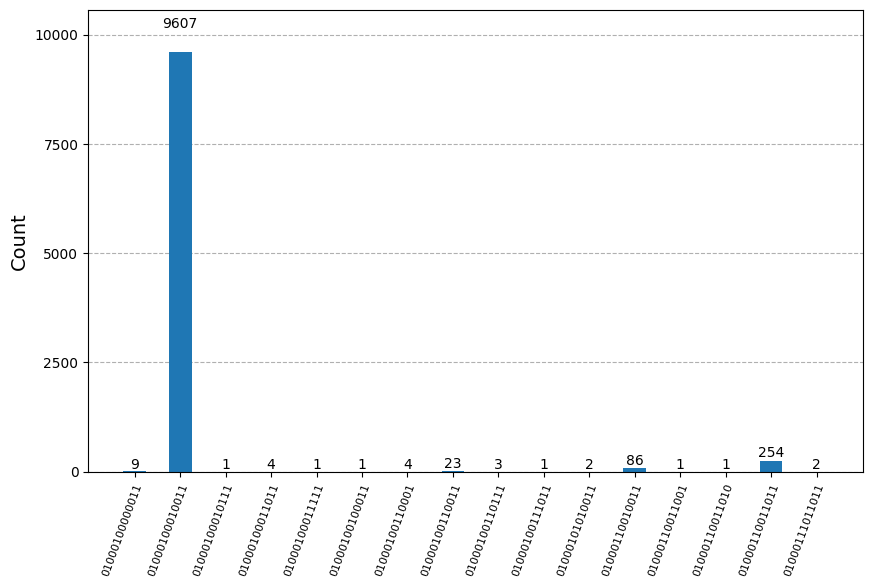}
        \caption{Probability distribution obtained from the CD-QAOA simulation with the MJ interaction at $l=2$.}
        \label{fig:cdqaoa_l2_pd_mj}
    \end{subfigure}

    \vspace{0.4cm}

    \begin{subfigure}{0.5\textwidth}
        \centering
        \includegraphics[width=\linewidth]{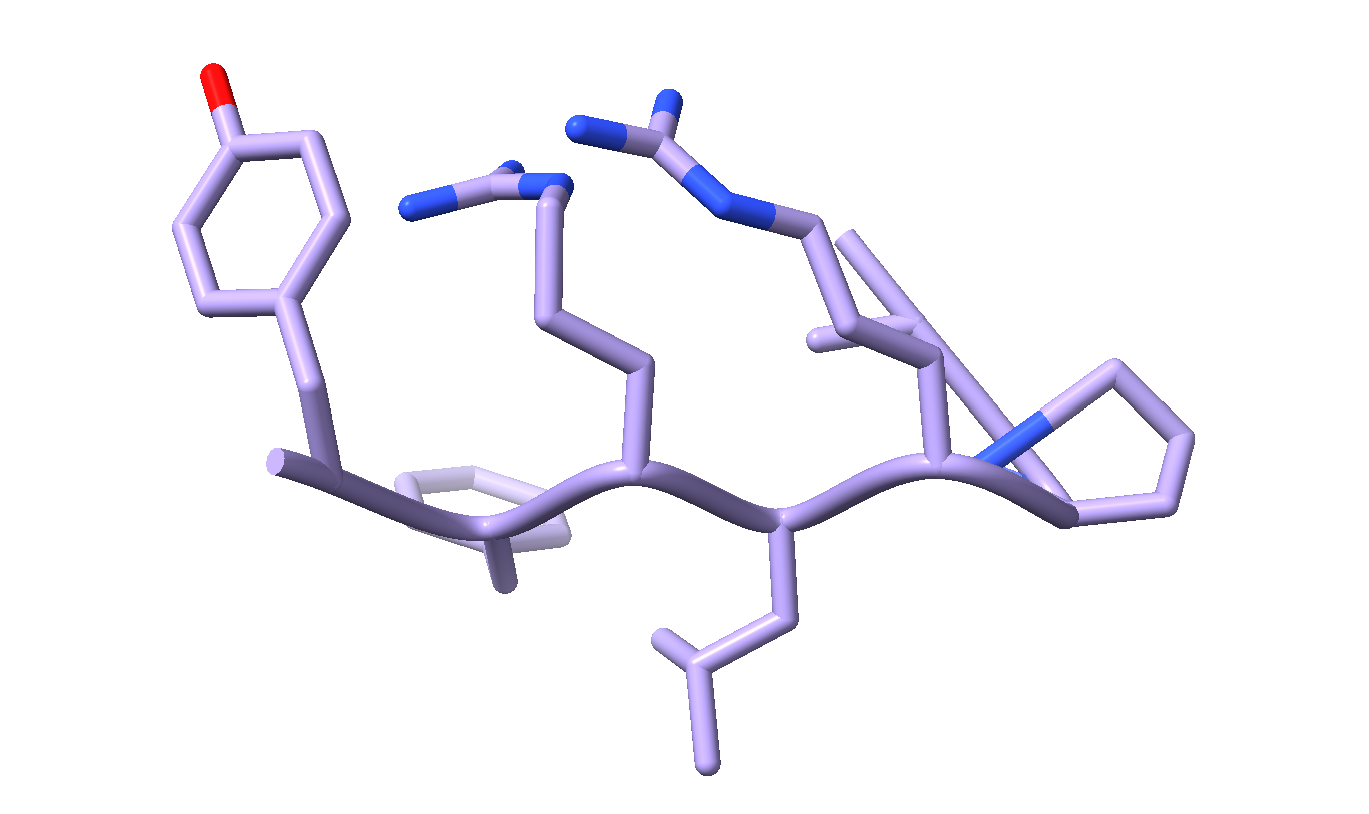}
        \caption{APRLRFY $C_{\alpha}$ structure generated from the bitstring 01001100110011.}
        \label{fig:apr_01001100110011}
    \end{subfigure}
    \hfill
    \begin{subfigure}{0.48\textwidth}
        \centering
        \includegraphics[width=\linewidth]{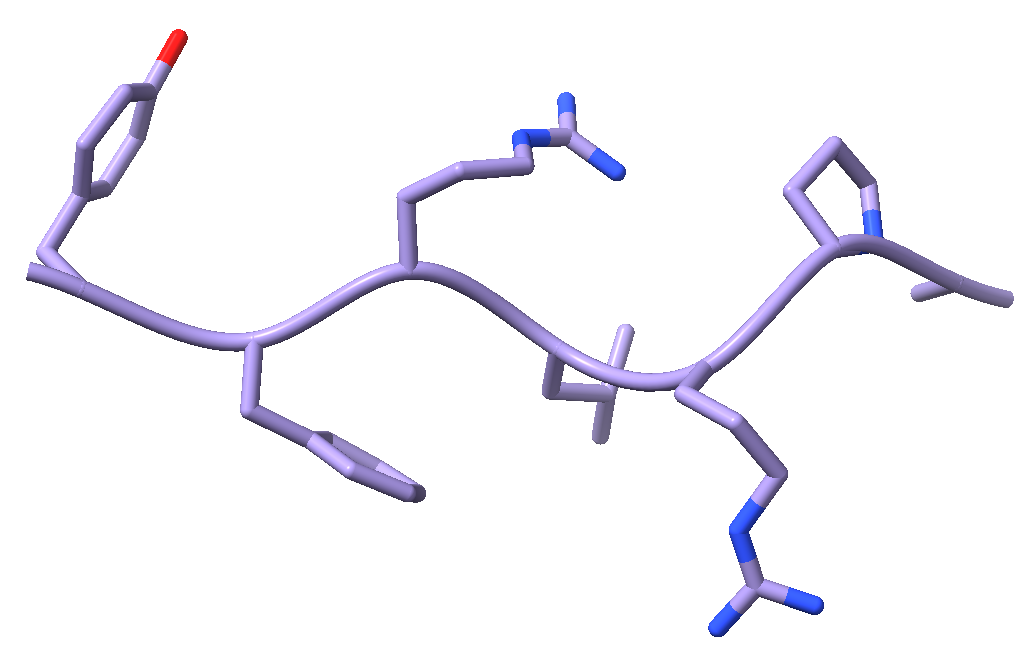}
        \caption{APRLRFY $C_{\alpha}$ structure generated from the bitstring 01001100110011 after a 10 ns MD simulation in water.}
        \label{fig:cd_mj_md}
    \end{subfigure}

    \caption{
    Probability distributions of CD-QAOA with the MJ interaction and reconstructed APRLRFY $C_{\alpha}$ structures.
    The optimal bitstring 01001100110011 was obtained from CD-QAOA with the MJ interaction at both $l=1$ and $l=2$.
    }
    \label{fig:mj_cdqaoa_aprlrfy_result}
\end{figure}
Here, we consider the case of the amino acid sequence APRLRFY ($N=7$), which requires 12 qubits and 2 interaction qubits. Following \cite{Robert}, the first four qubits can be fixed, and following \cite{Chandarana2022}, the 6th qubit can also be fixed.
\begin{equation}
[01][00][q_0 1] [q_1 q_2] \cdots \nonumber
\end{equation}

We implemented quantum circuits for the $l=1$ and $l=2$ cases. Using the COBYLA optimizer in Qiskit, we trained the quantum circuit to minimize $H_{gc}+H_{int}$. Fig.~\ref{fig:cdqaoa_l1_pd} shows the probability distribution obtained from the trained circuit for the $l=1$ case. The three most frequently counted sequences were 010011100010(00) and 010011100011(00 or 01), where the interaction qubits are indicated in parentheses. Fig.~\ref{fig:cdqaoa_l2_pd} shows the result for the $l=2$ case. The most frequently counted sequence was 010011100011(01). Therefore, 010011100011 was identified as the common optimal sequence. The interaction qubit can be set to 0 to make the interaction term 0. However, when an attractive interaction exists, the interaction term $\epsilon_{ij}$ should be included. The second interaction qubit corresponds to the interaction between proline (P) and tyrosine (Y), and according to \cite{miyazawa1996}, this pair exhibits an attractive interaction.\\

We reconstructed the protein structure based on the $C_{\alpha}$ structure. The length of each edge in the lattice was set to $3.8~\text{\AA}$ and converted into XYZ coordinates to generate a new PDB file. Following \cite{rotkiewicz2008pulchra}, all atoms were reconstructed. Fig.~\ref{fig:cdqaoa_aprlrfy_result} shows the reconstructed structure. In Fig.~\ref{fig:apr_010011100010}, where all interaction qubits are 0. Compared with the structure shown in Fig.~\ref{fig:apr_010011100010}, the reconstructed structure in Fig.~\ref{fig:apr_010011100011} shows a more compact conformation, with Pro2 positioned closer to Tyr7. This result suggests that the bitstring 010011100011 reflects increased residue--residue proximity between Pro2 and Tyr7 and corresponds to a more compact conformational state of the APRLRFY peptide.\\

Unlike the first CD-QAOA approach, in which interaction terms were introduced only between the two specific residues P and Y in the heptapeptide, we replaced all the interaction terms with the Miyazawa--Jernigan matrix and investigated peptide folding behavior. When the Miyazawa--Jernigan matrix was employed as the interaction term, the most probable peptide conformation differed from the previously obtained structure and instead adopted an extended configuration, shown in purple in Fig.~\ref{fig:apr_01001100110011}.\\

Fig.~\ref{fig:cdqaoa_l1_pd_mj} and Fig.~\ref{fig:cdqaoa_l2_pd_mj} show the probability distributions obtained from the CD-QAOA simulations with the MJ interaction at $l=1$ and $l=2$, respectively. The bitstring 01001100110011 was obtained as the dominant structural candidate in both cases. The corresponding reconstructed APRLRFY $C_{\alpha}$ structure is shown in Fig.~\ref{fig:apr_01001100110011}. To evaluate the structural stability of the CD-QAOA-derived conformation in an aqueous environment, a 10 ns MD simulation was performed in water. As shown in Fig.~\ref{fig:cd_mj_md}, the peptide adopted a more extended conformation after MD simulation compared with the initial reconstructed structure. This structural change suggests that the CD-QAOA-derived lattice structure was further relaxed under the atomistic force field and solvent environment, resulting in changes in the backbone and side-chain conformations.

\subsection{Hartree Fock Calculation}
\begin{figure}[H]
    \centering

    \begin{subfigure}{0.48\textwidth}
        \centering
        \includegraphics[width=\linewidth]{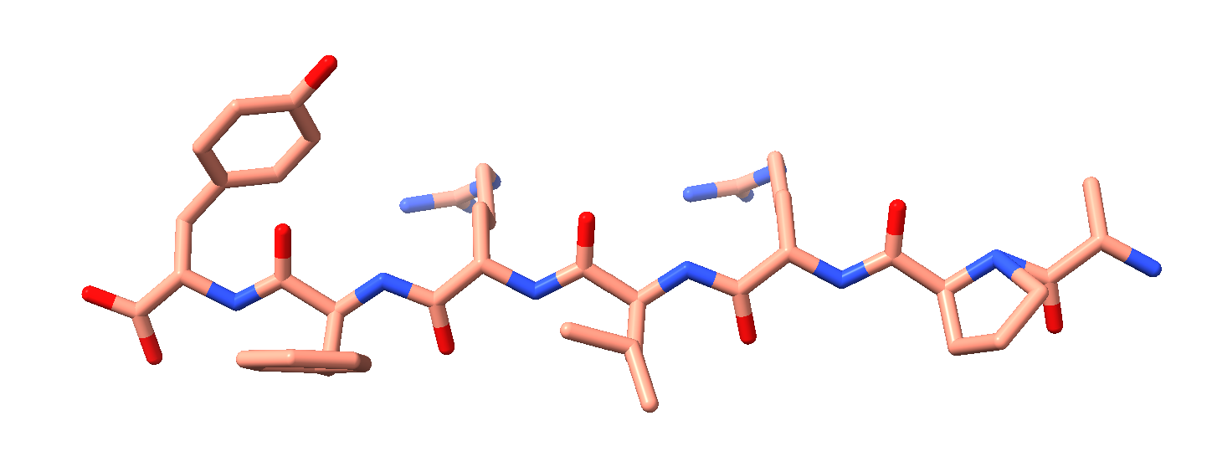}
        \caption{Initial configuration}
        \label{fig:hf_initial}
    \end{subfigure}
    \hfill
    \begin{subfigure}{0.45\textwidth}
        \centering
        \includegraphics[width=\linewidth]{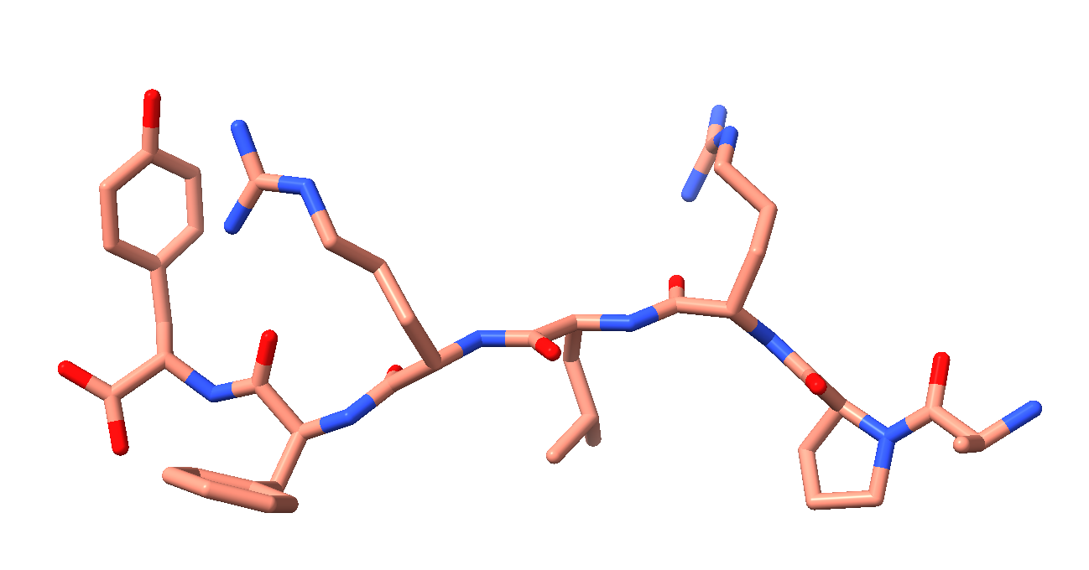}
        \caption{Hartree--Fock ground-state configuration of APRLRFY}
        \label{fig:hf_final}
    \end{subfigure}

    \caption{Structural comparison of the initial and Hartree--Fock-optimized ground-state configurations of the APRLRFY peptide.}
    \label{fig:hf_result}
\end{figure}

At the UHF/6-31G level, the electronic structure optimization of the heptapeptide APRLRFY converged to a total system energy of -3085.72136976 hartree. The fully extended conformation, used as the initial structure, underwent structural changes during the optimization process due to intramolecular interactions. Unlike the initial linear conformation, the optimized structure adopted a spatially rearranged conformation.\\

This conformational change arises from thermodynamic stabilization, which minimizes electrostatic repulsion between the two positively charged arginine (Arg) residues and reduces steric hindrance involving the aromatic phenylalanine (Phe) and tyrosine (Tyr) residues. Consequently, it was confirmed that the APRLRFY peptide forms its characteristic three-dimensional conformation in a gas-phase environment through inter-residue charge repulsion and van der Waals interactions.

\subsection{DFT Calculation}
\begin{figure}[h]
    \centering
    \includegraphics[width=0.6\textwidth]{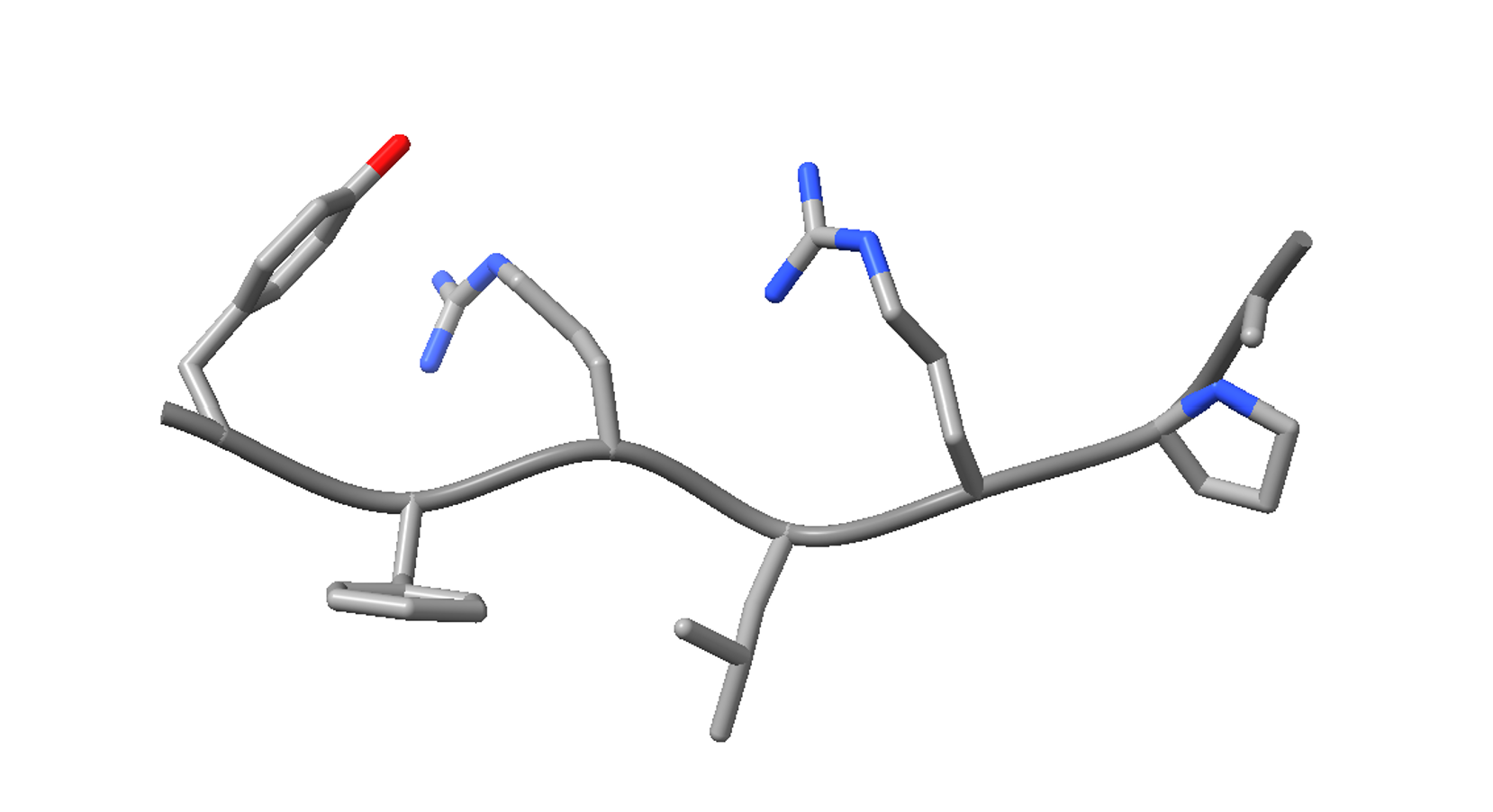}
    \caption{
    DFT-optimized structure of the APRLRFY peptide in an aqueous environment using the SMD implicit solvation model at the B3LYP/6-31G(d) level of theory.
    }
    \label{fig:dft_smd_structure}
\end{figure}

The DFT-SMD optimization of the APRLRFY peptide converged to a total system energy of -3107.33992520 hartree. The optimized structure exhibited a relatively extended backbone conformation. As shown in Fig.~\ref{fig:dft_smd_structure}, several side chains were oriented outward from the peptide backbone, while the overall peptide structure maintained a non-compact conformation. This structural feature may reflect the combined effects of intramolecular interactions and solvent stabilization under the implicit water environment.
Frequency analysis was subsequently performed at the same level of theory to verify whether the optimized structure corresponds to an energy minimum. No imaginary frequencies were observed, indicating that the optimized structure is not a transition state but a stable local minimum on the potential energy surface. 

\subsection{MD simulation}

\begin{figure}[h]
    \centering
    \includegraphics[width=0.7\textwidth]{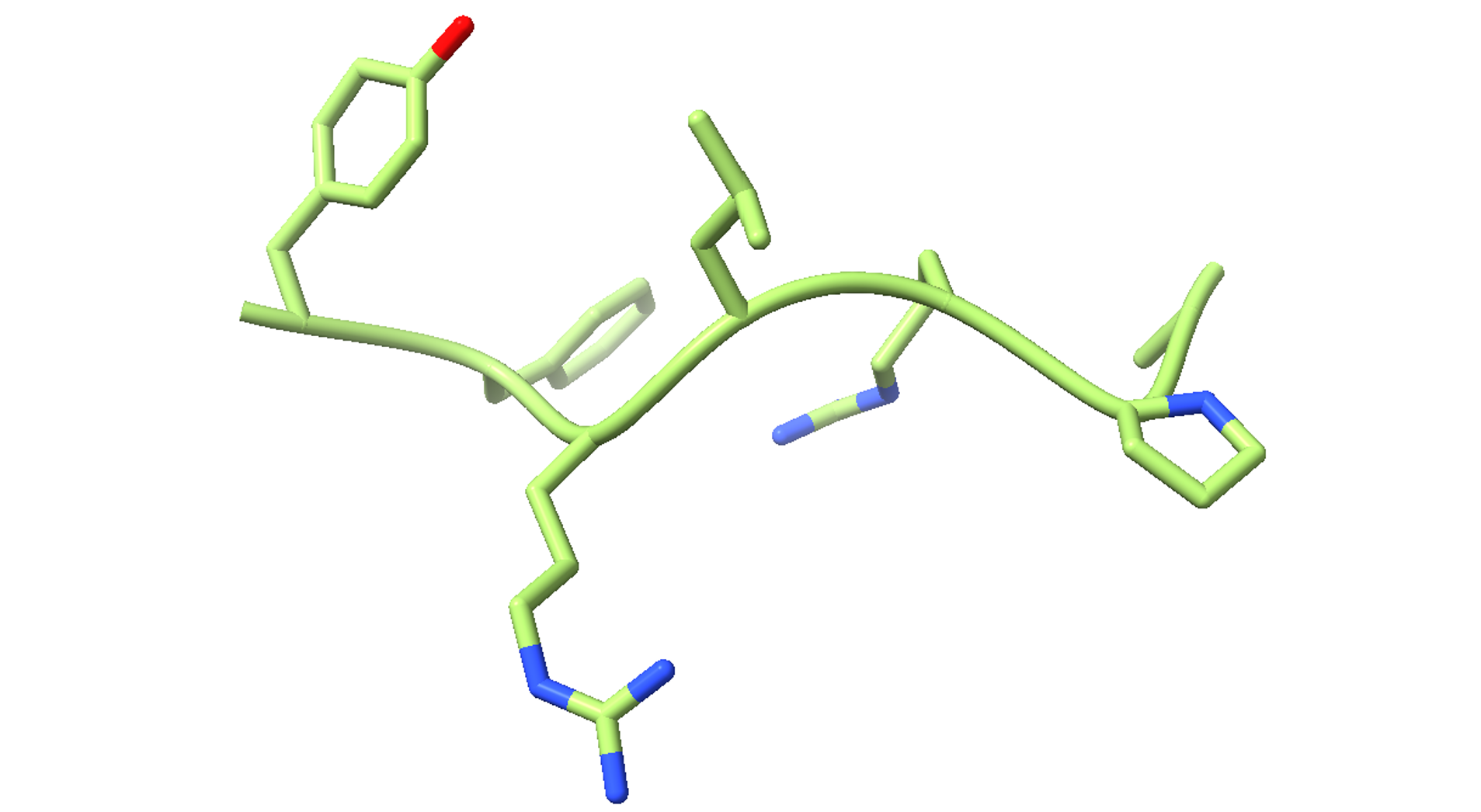}
    \caption{ 
    MD-refined structure of the APRLRFY peptide after relaxation of the AI-based initial structure.}
    \label{fig:ai_md}
\end{figure}

The three-dimensional structure of a heptapeptide composed of seven amino acids was initially predicted using AlphaFold3. As shown in Fig.~\ref{fig:ai_md}, the predicted model exhibited a partially folded structure. However, a detailed examination of the downloaded atomic coordinates revealed numerous steric clashes, particularly between the side-chain atoms (not shown). The sequence contains two arginine residues with bulky, positively charged side chains, and near the C-terminus, phenylalanine and tyrosine residues with large aromatic rings are present. The presence of these sterically large functional groups makes the predicted structure sensitive to even small geometric errors, thereby increasing the likelihood of steric conflicts in the initial model. A 10 ns MD simulation was performed in explicit water environment. The minimized structure exhibited a relaxed extended conformation.

\subsection{H-REMD}
\begin{figure}[h]
\centering
\includegraphics[width=1.0\linewidth]{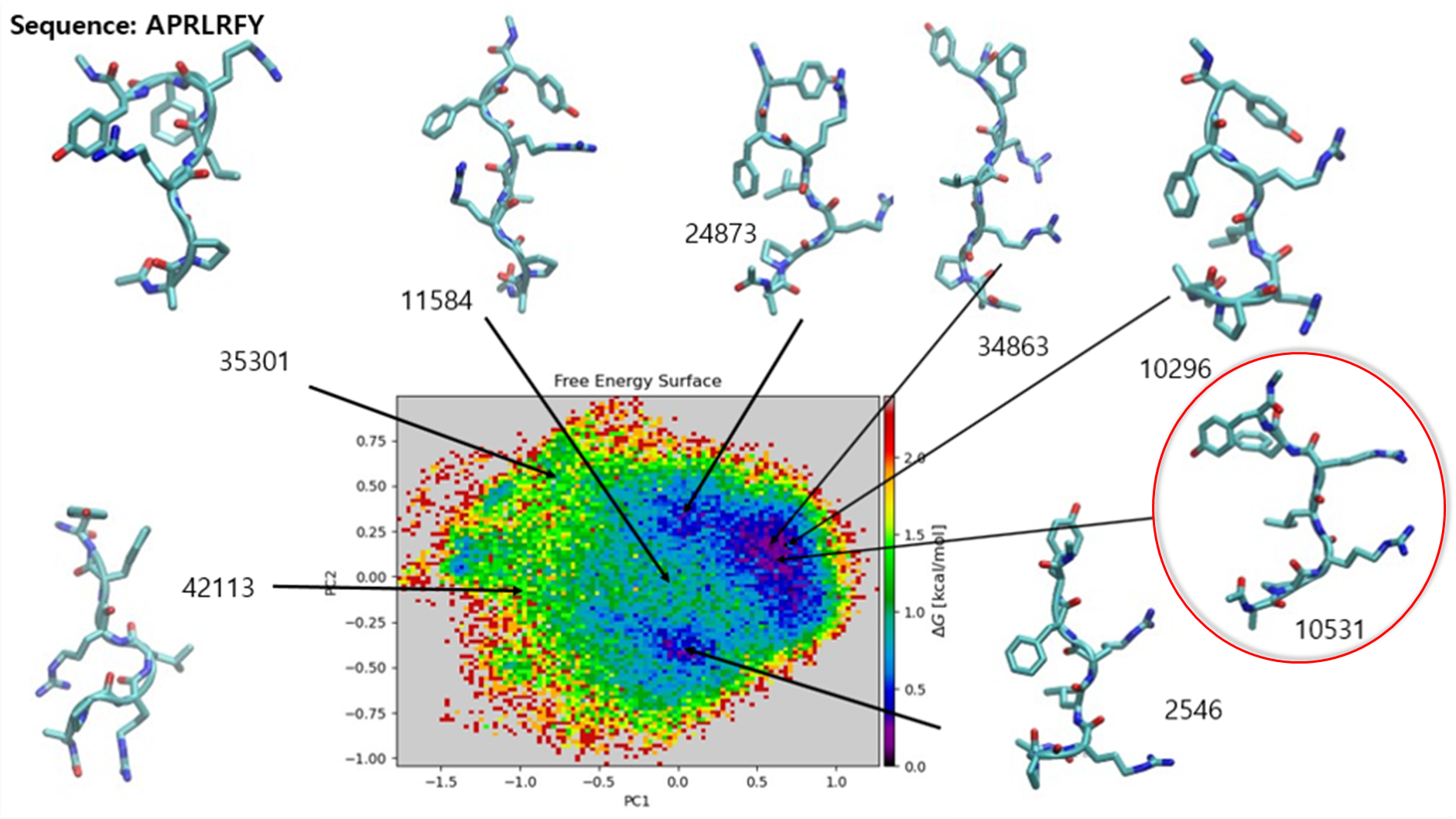}
\caption{\label{fig:hermd} Conformational space sampling of APRLRFY by H-REMD calculation.}
\end{figure}
To further explore the conformational landscape of the heptapeptide, we applied the H-REMD sampling method. Among the peptide conformations corresponding to local minima, the global minimum structure with the lowest free energy (structure 10531, shown in the red circle in Fig.~\ref{fig:hermd}) also displayed a stretched configuration.\\

\subsection{Backbone RMSD Analysis}

\begin{figure}[H]
\centering
\includegraphics[width=1.0\linewidth]{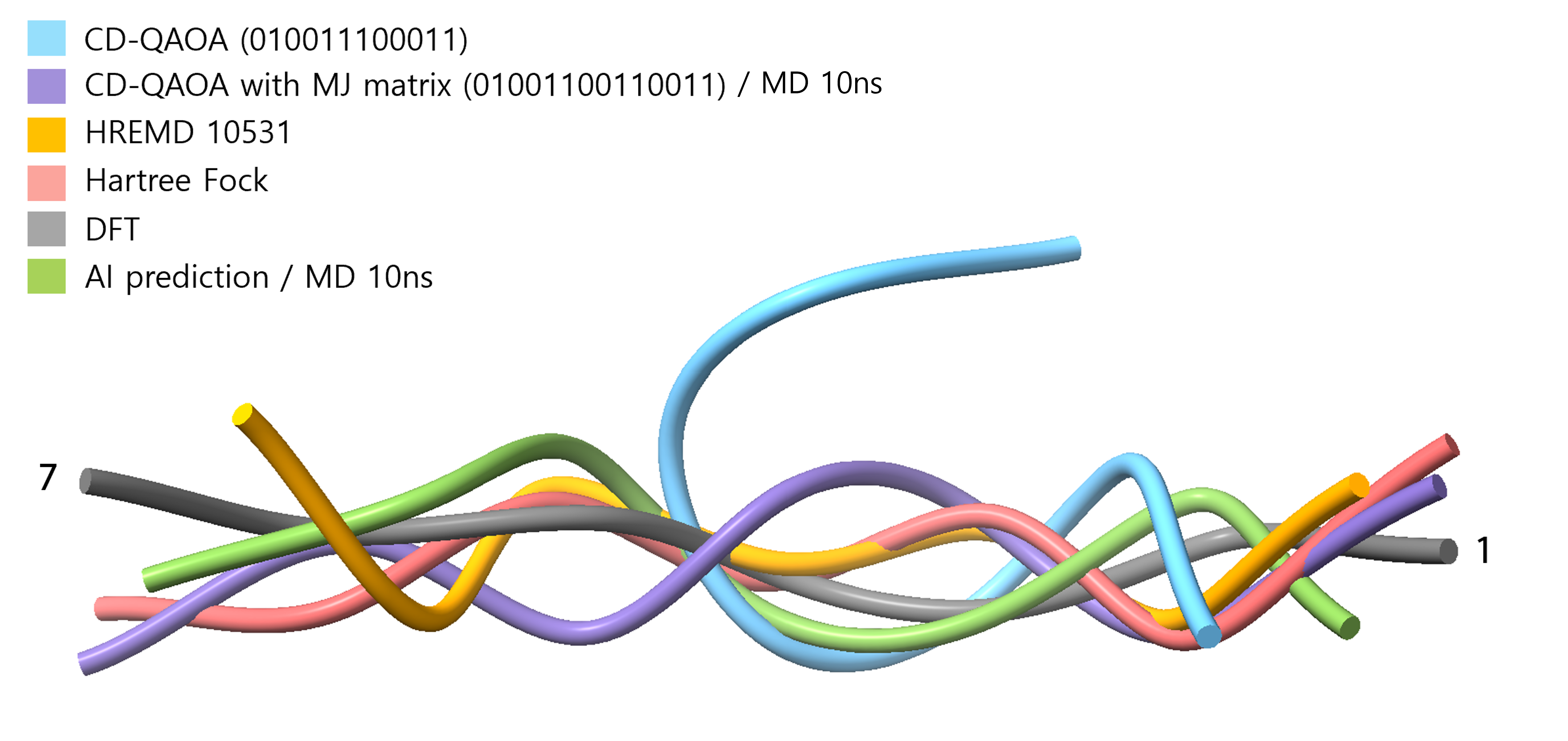}
\caption{\label{fig:Fig5} Alignments of predicted structures of APRLRFY using different methods. 1 and 7 denote N-terminus and C-terminus of the peptide.}
\label{fig:conclustion_fig}
\end{figure}

\begin{table}[H]
\centering
\caption{Pairwise RMSD values between predicted APRLRFY peptide structures.}
\label{tab:rmsd_comparison}

{\fontsize{9pt}{12pt}\selectfont
\renewcommand{\arraystretch}{1.25}
\setlength{\tabcolsep}{4pt}
\begin{tabularx}{\linewidth}{>{\centering\arraybackslash}p{0.25\linewidth}
                                >{\centering\arraybackslash}X
                                >{\centering\arraybackslash}X
                                >{\centering\arraybackslash}X
                                >{\centering\arraybackslash}X
                                >{\centering\arraybackslash}X}
\hline
\multicolumn{6}{c}{\textbf{RMSD (\AA)}} \\
\hline
\textbf{Structure} & \textbf{HF} & \textbf{DFT} & \textbf{MD} & \textbf{H-REMD} & \textbf{CD-QAOA} \\
\hline
Hartree--Fock & 0.00 & 1.67 & 1.63 & 2.20 & 1.39 \\
DFT & 1.67 & 0.00 & 1.42 & 2.14 & 1.90 \\
MD & 1.63 & 1.42 & 0.00 & 2.22 & 1.70 \\
H-REMD & 2.20 & 2.14 & 2.22 & 0.00 & 2.37 \\
CD-QAOA (with MJ)/MD & 1.39 & 1.90 & 1.70 & 2.37 & 0.00 \\
\hline
\end{tabularx}
}

\end{table}

The CD-QAOA/MD peptide structure showed overall low backbone root mean square deviation (RMSD) values compared with the HF (1.39~\AA), DFT (1.90~\AA), and MD (1.70~\AA) structures. In particular, the RMSD value relative to the HF structure was the lowest (1.39~\AA), indicating that the CD-QAOA/MD structure exhibited the highest structural similarity to the HF-optimized structure. In contrast, the RMSD relative to the H-REMD structure was slightly higher, at 2.37~\AA.\\

The RMSD between the CD-QAOA/MD and MD structures was also low, at 1.70~\AA, indicating that the CD-QAOA/MD structure retains structural features similar to those of the MD-generated structure. In contrast, the RMSD relative to the H-REMD structure was the largest (2.37~\AA), implying that that H-REMD sampled a distinct conformational basin that may be closer to the global minimum on the peptide free-energy landscape. In contrast, the CD-QAOA/MD and HF structures are likely to represent local minimum conformations obtained under their respective computational conditions. Their relatively low RMSD value indicates that the two approaches converged to structurally related local minima, despite differences in the underlying optimization methodologies.\\

Overall, these results indicate that CD-QAOA /MD (01001100110011) can predict a peptide structure comparable to those obtained from conventional quantum chemical and MD-based approaches.

\subsection{Side-Chain Conformation Analysis}
\begin{figure}[H]
\centering
\includegraphics[width=1.0\linewidth]{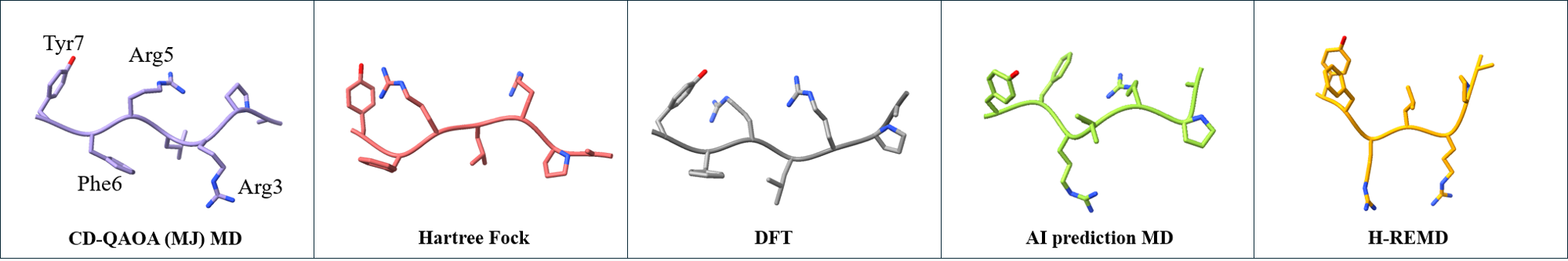}
\caption{\label{fig:aa_all} Side-chain orientations in APRLRFY peptide structures predicted by
different computational methods.}
\label{fig:aa_all}
\end{figure}
The orientations of the Arg3 and Arg5 side chains in the APRLRFY peptide were found to depend on the computational method employed. In the CD-QAOA/MD and MD structures, the Arg3 and Arg5 side chains adopted opposite orientations, whereas in the HF, DFT, and H-REMD structures they were aligned in the same direction. These differences likely reflect method-dependent variations in the balance among Arg–Arg electrostatic repulsion, backbone conformational constraints, terminal capping effects (in the case of H-REMD), and Arg–aromatic interactions involving the Phe6 and Tyr7 residues.\\

The opposite orientations of Arg3 and Arg5 observed in the CD-QAOA/MD and MD structures may represent a favorable arrangement for reducing electrostatic repulsion between the positively charged guanidinium groups of the two arginine residues. During MD relaxation, the conformational landscape is influenced by force-field-derived backbone dihedral potentials, steric interactions, and chirality-dependent constraints associated with L-amino acids. Consequently, the observed Arg3–Arg5 arrangement is likely the result of a cooperative adjustment of backbone torsions and side-chain packing that minimizes the overall potential energy.\\

In contrast, although H-REMD is also based on molecular dynamics simulations, Arg3 and Arg5 adopted the same orientation. This difference may arise from the N-terminal acetylation and C-terminal N-methyl amidation employed in the H-REMD model. These terminal modifications alter the charge distribution and conformational flexibility of the peptide backbone, thereby influencing the packing of neighboring side chains and potentially favoring a distinct local minimum conformation.\\

In the HF and DFT structures, the same-direction arrangement of Arg3 and Arg5 may reflect the direct consideration of electrostatic and non-covalent interactions within the peptide during electronic-structure-based calculations. In particular, possible cation--$\pi$ interactions \cite{Gallivan1999} between Arg5 and the aromatic rings of the adjacent Phe6/Tyr7 residues may partially compensate for Arg--Arg electrostatic repulsion, thereby stabilizing the same-direction side-chain arrangement. \\

Overall, the difference in Arg3--Arg5 orientation in APRLRFY can be viewed as the result of varying relative contributions from electrostatic repulsion, chirality-dependent backbone constraints, terminal capping effects, and Arg--aromatic interactions depending on the calculation method.

\section{Conclusion}
In this study, we optimized a CD-QAOA-based quantum algorithm to fold a seven-residue peptide, APRLRFY, on a tetrahedral lattice. When interactions were modeled using the Miyazawa--Jernigan matrix, the resulting structures showed strong agreement with those obtained from HF-based quantum chemical calculations, as well as with molecular dynamics and H-REMD simulations that capture peptide behavior in aqueous solution, despite some dependence on the chosen parameters.\\

We performed molecular dynamics (MD) simulations in explicit water using peptide structures generated by two CD-QAOA models: (i) a model incorporating an additional interaction between Pro2 and Tyr7 (bitstring: 010011100011) and (ii) the MJ interaction model (bitstring: 0100110011001).\\

For the model with the additional Pro2–Tyr7 interaction (bitstring: 010011100011), the initially folded structure rapidly relaxed into an extended conformation during the MD simulation. This observation suggests that the fixed interaction imposed between Pro2 and Tyr7 may have artificially stabilized the folded state, leading to an overestimation of its thermodynamic stability (data not shown).\\

In contrast, the structure generated by the MJ interaction model (bitstring: 0100110011001) retained its initially extended conformation throughout the simulation, with no significant structural changes observed in explicit water. This result indicates that the extended state is thermodynamically stable under the simulation conditions. \\

In the present study, only the $H_{gc}+H_{int}$ interaction terms were considered. However, a practically applicable CD-QAOA framework should also account for peptide chirality, conformational flexibility, and the dynamic behavior of peptides in aqueous environments. A quantum–classical hybrid approach that integrates CD-QAOA ($H_{gc}+H_{int}$) with molecular dynamics (MD) simulations offers several advantages in this regard. In particular, MD force fields inherently preserve peptide chirality through stereochemical constraints while simultaneously capturing conformational fluctuations and solvent-dependent dynamics in aqueous solution. These features provide physically realistic structural information that remains difficult to obtain directly using current quantum algorithms for short peptide structure prediction.\\

For longer peptides and intrinsically disordered proteins (IDPs), which are of greater practical relevance, a hybrid strategy integrating CD-QAOA with classical MD simulation techniques is expected to provide both improved computational efficiency and enhanced structural prediction accuracy.\\

\section{Acknowledgements}
This work was supported by the National Research Foundation of KOREA (NRF) (RS-2023-00257994).

\bibliographystyle{unsrt}
\bibliography{ref}

\end{document}